\documentclass[aps,prd,superscriptaddress,showpacs,preprint,amsmath,amssymb]{revtex4}
\usepackage{graphicx, bm}
\usepackage[usenames]{color}
\usepackage{float}

\begin{document}

%\tightenlines
\draft
\title{Model-independent bounds probing on the electromagnetic dipole moments of the $\tau$-lepton at the CLIC}

\author{M. K\"{o}ksal\footnote{mkoksal@cumhuriyet.edu.tr}}
\affiliation{\small Deparment of Optical Engineering, Cumhuriyet University, 58140, Sivas, Turkey.\\}

\author{A. A. Billur\footnote{abillur@cumhuriyet.edu.tr}}
\affiliation{\small Deparment of Physics, Cumhuriyet University, 58140, Sivas, Turkey.\\}

\author{ A. Guti\'errez-Rodr\'{\i}guez\footnote{alexgu@fisica.uaz.edu.mx}}
\affiliation{\small Facultad de F\'{\i}sica, Universidad Aut\'onoma de Zacatecas\\
         Apartado Postal C-580, 98060 Zacatecas, M\'exico.\\}

\author{ M. A. Hern\'andez-Ru\'{\i}z\footnote{mahernan@uaz.edu.mx}}
\affiliation{\small Unidad Acad\'emica de Ciencias Qu\'{\i}micas, Universidad Aut\'onoma de Zacatecas\\
         Apartado Postal C-585, 98060 Zacatecas, M\'exico.\\}

\date{\today}
%\maketitle

\begin{abstract}
% insert abstract here

We establish model independent bounds on the anomalous magnetic and electric dipole moments of the tau-lepton using the two-photon
processes $\gamma\gamma \to \tau^+\tau^-$ and $\gamma\gamma \to \tau^+\tau^-\gamma$. We use ${\cal L}=10, 50, 100, 300, 500, 1000, 1500,
2000, 3000 \hspace{0.8mm}fb^{-1}$ of data collected with the future $e^+e^-$ linear collider such as the CLIC at $\sqrt{s}=380, 1500, 3000
\hspace{0.8mm}GeV$ and we consider systematic uncertainties of $\delta_{sys}=0, 3, 5\hspace{1mm}\%$. Precise bounds at $95\%$ C. L. on
the anomalous dipole moments to the tau-lepton $-0.00015\leq \tilde{a}_\tau \leq 0.00017$ and $|\tilde{d}_\tau(ecm)|=9.040\times 10^{-19}$
are set from our study. Our results show that the processes under consideration are a very good prospect for probing the dipole moments of
the tau-lepton at the future $e^+e^-$ linear collider at the $\gamma\gamma$ mode.

\end{abstract}


\vspace{5mm}

\maketitle

\section{Introduction}

One of the greatest achievement of the Standard Model (SM)  \cite{Glashow,Weinberg,Salam}  is the measurement of the electric (EDM) and
magnetic (MM) dipole moments of the electron and muon $g-2$, which have been measured with a excellent precision of

\begin{equation}
a^{Exp}_e= 1 159 652 180.73 (28)\times 10^{-12} \hspace{1cm} \mbox{[0.24\hspace{0.8mm}ppb]} \hspace{5mm} \mbox{\cite{Hanneke}},
\end{equation}

\begin{equation}
a^{Exp}_\mu= 11659209.1(5.4)(3.3)\times 10^{-10}  \hspace{1cm} \mbox{[0.54\hspace{0.8mm}ppm]} \hspace{5mm} \mbox{\cite{Bennett}},
\end{equation}

\noindent respectively, and the theoretical prediction of the SM \cite{Data2016} is given by

\begin{equation}
a^{SM}_\mu= 116591803(1)(42)(26) \times 10^{-11}.
\end{equation}

On the other hand, in comparison with the electron or muon mass, the tau-lepton has a large mass of $m_\tau=1776.82\pm 0.16\hspace{0.8mm}MeV $
\cite{Data2016}, allowing one to expected an essential enhancement in the sensitivity to the effects of new physics beyond the Standard Model (BSM),
such as its dipole moments \cite{Passera1}. However the very short lifetime of this unstable particle makes it impossible to directly
measure their electromagnetic properties. Therefore, indirect information must be obtained by precisely measuring cross sections and decay rates
in processes involving the emission of a real photon by the tau-lepton.

With respect to the anomalous magnetic moment of the $\tau$-lepton, the SM prediction is $a^{SM}_{\tau} = 117721(5)\times 10^{-8}$
\cite{Samuel,Hamzeh} and the respective EDM $d_\tau$ is generated by the GIM mechanism only at very high order
in the coupling constant \cite{Barr}. The error with an order of magnitude of $10^{-8}$ is an indication that SM extensions predicting
values for $a_\tau$ above this level they are worth studying, as well as it is worthwhile to study new mechanisms and new modes of
production of tau pairs in association with a photon $\tau^+\tau^-\gamma$ at the future $e^+e^-$ linear collider at the $\gamma\gamma$ mode.

The SM predicts CP violation, which is necessary for the existence of the EDM of a variety physical
systems. The EDM provides a direct experimental probe of CP violation \cite{Christenson,Abe,Aaij}, a feature of the SM and BSM physics.
Precise measurement of the EDM of fundamental charged particles provides a significant probe of physics BSM.

\begin{table}[!ht]
\caption{Summary of experimental and theoretical limits on the electromagnetic dipole moments of the tau-lepton.}
\begin{center}
\begin{tabular}{|c| c| c| c|}
\hline
{\bf Collaboration}  &    {\bf Experimental limit}            & {\bf C. L.}  &  {\bf Reference}\\
\hline\hline
DELPHI       &   $-0.052 < a_\tau < 0.013$                    & $95 \%$  & \cite{Abdallah} \\
\hline
L3           &   $-0.052 < a_\tau < 0.058$                    & $95 \%$  & \cite{Acciarri} \\
\hline
OPAL         &   $-0.068 < a_\tau < 0.065$                    & $95 \%$ & \cite{Ackerstaff} \\
\hline
\hline
{\bf Collaboration}                       &     {\bf Experimental limit}                             &    {\bf C. L.}  &  {\bf Reference}\\
\hline
\hline
BELLE       &     $-2.2 < Re(d_\tau(10^{-17}e cm)) < 4.5$   &   $95 \%$  & \cite{Inami} \\
            &     $-2.5 < Im(d_\tau(10^{-17}e cm)) < 0.8$   &   $95 \%$  &               \\
\hline
DELPHI      &     $-0.22 < d_\tau(10^{-16}e cm) < 0.45$     &   $95 \%$  & \cite{Abdallah} \\
\hline
L3          &     $ |Re(d_\tau(10^{-16}e cm))| < 3.1$       &   $95 \%$  & \cite{Acciarri} \\
\hline
OPAL        &     $ |Re(d_\tau(10^{-16}e cm))| < 3.7$       &   $95 \%$  & \cite{Ackerstaff} \\
\hline
ARGUS       &     $ |Re(d_\tau(10^{-16}e cm))| < 4.6$       &   $95 \%$  & \cite{Albrecht} \\
            &     $ |Im(d_\tau(10^{-16}e cm))| < 1.8$       &   $95 \%$  &                  \\
\hline\hline
{\bf Model} &     {\bf Theoretical limit}                   &    {\bf C. L.}  &  {\bf Reference}\\
\hline\hline
L3 data      &     $a_\tau \leq 0.11$                       &    $90 \%$   & \cite{Grifols} \\
\hline
Electroweak Measurements &     $-0.004 < a_\tau < 0.006$    &    $95 \%$  & \cite{Escribano} \\
\hline
LEP1, SLC, LEP2 Data     &     $-0.007 < a_\tau < 0.005$    &    $95 \%$  & \cite{Sprinberg} \\
\hline
Total cross section      &     $ a_\tau < 0.023$            &    $95 \%$  & \cite{Silverman} \\
\hline\hline
{\bf Model} &     {\bf Theoretical limit}                   &    {\bf C. L.}  &  {\bf Reference}\\
\hline\hline
L3 data      &     $d_\tau \leq 6\times 10^{-16} e cm$      &    $90 \%$   & \cite{Grifols} \\
\hline
Electroweak Measurements &   $d_\tau \leq 1.1\times 10^{-17} e cm$         &    $95 \%$  & \cite{Escribano} \\
\hline
Cross section &   $d_\tau \leq 1.6\times 10^{-16} e cm$      &    $90 \%$  & \cite{DelAguila} \\
\hline
\hline
\end{tabular}
\end{center}
\end{table}

The sensitivity to the MM and EDM of the tau-lepton has been studied in different context, both theoretical and experimental and some of which
are summarized in Table I. Furthermore, there is an extensive theoretical work done in models BSM that contribute to dipole moments of charged
leptons: Extra dimensions \cite{Iltan1},  Seesaw model \cite{Dutta}, version III of the 2HDM \cite{Iltan2}, Non-commutative geometry \cite{Iltan3}, Non-universal extra dimensions \cite{Iltan4}, Left-Right symmetric model \cite{Gutierrez1}, $E_6$ Superstring models \cite{Gutierrez2}, Simplest
little Higgs model \cite{Gutierrez3}, 331 model \cite{Gutierrez4}. There are also bounds independent of the model such as $\gamma p$ collisions
\cite{Koksal1}, $e^-\gamma$ scattering \cite{Ozguven} and $\gamma\gamma$ collisions \cite{Billur,Sampayo}. Other limits on the MM and EDM of the
$\tau$-lepton are reported in Refs. \cite{Passera1,Eidelman1,Galon,Arroyo1,Arroyo2,Xin,Pich,Atag1,Lucas,Gutierrez5,Passera2,Passera3,Bernabeu,Gutierrez6,Bernreuther}.

In this paper, using $\gamma\gamma \to \tau^+\tau^-$ and $\gamma\gamma \to \tau^+\tau^-\gamma$ reactions, we establish
model independent limits on the dipole moments of the tau-lepton, and we improve the existing bounds on $a_\tau$ and $d_\tau$.
An interesting feature of these reactions is that they are extremely clean process because it has not interference with weak
interactions, being a purely process of Quantum Electrodynamics (QED). Furthermore, the high center-of-mass energies proposed
for the Compact Linear Collider (CLIC) make of it an appropriate machine to probing the MM and EDM which are more sensitive
with the high energy and high luminosity of the collider.

The CLIC \cite{Linssen,Accomando,Abramowicz,Dannheim} is a proposal for a future $e^+e^-$ linear collider
at CERN in the High Luminosity-Large Hadron Collider (HL-LHC) era. The machine is designed to make full use of the physics potential
of CLIC with initial operation at center-of-mass energy of $\sqrt{s}=380\hspace{0.8mm}GeV$ and luminosity of ${\cal L}=500\hspace{0.8mm}fb^{-1}$.
The following stages at center-of-mass energies of $\sqrt{s}=1500\hspace{0.8mm}GeV$ $({\cal L}=1500 fb^{-1}$) and $\sqrt{s}=3000\hspace{0.8mm}GeV$
$({\cal L}=3000\hspace{0.8mm}fb^{-1}$) focus on exploring physics BSM. In summary, the CLIC project offers a rich physics program for about 20 years,
with discovery potential to new physics, that can reach scales of up to several tens of TeV, through indirect searches with precision measurements.

For our study, we consider the following parameters of the CLIC: $\sqrt{s}=380, 1500, 3000\hspace{0.8mm}GeV$,
${\cal L}=10, 50, 100, 300, 500, 1000, 1500, 2000, 3000 \hspace{0.8mm}fb^{-1}$ and we consider systematic
uncertainties of $\delta_{sys}=0, 3, 5\hspace{1mm}\%$, with these parameters as input, we established model
independent bounds on the electromagnetic dipole moments of the $\tau$-lepton at $95\%$ C.L.. We get stringent
limits in comparison with the bounds obtained by the DELPHI, L3, OPAL, BELLE and ARGUS Collaborations \cite{Abdallah,Acciarri,Ackerstaff,Inami,Albrecht}
(see Table I).

The remainder of the paper is organized as follows: In Section II, we study the total cross section and the electromagnetic
dipole moments of the tau-lepton through the $\gamma\gamma \to \tau^+\tau^-$ and $\gamma\gamma \to \tau^+\tau^-\gamma$ reactions.
Section III provides the conclusions.

\section{Two-photon processes $\gamma\gamma \to \tau^+\tau^-$  and $\gamma\gamma \to \tau^+\tau^-\gamma$}

For our study, we will take advantage of our previous works on the collision modes $\gamma\gamma$, $\gamma\gamma^*$ and $\gamma^*\gamma^*$ \cite{Billur,Gutierrez7,Koksal2,Koksal3} for calculate the total cross section for the $\gamma\gamma \to \tau^+\tau^-$ and $\gamma\gamma \to \tau^+\tau^-\gamma$ reactions. The corresponding Feynman diagrams for these processes are given in Figs. 1 and 2, respectively.

In our study we deduce bounds on the electromagnetic dipole moments of the $\tau$-lepton $a_\tau$ and $d_\tau$ via the two-photon processes
$\gamma\gamma \to \tau^+\tau^-$ and $\gamma\gamma \to \tau^+\tau^-\gamma$. These processes are of interest for a number of reasons: First,
are sensitive to the $a_\tau$ and $d_\tau$. Additionally, increased cross sections for high energies and the absence or stong suppression
of weak contributions are further complementary aspects of the two-photon processes in the contrast with the direct processes $e^+e^- \to \tau^+\tau^-$ \cite{DelAguila,Albrecht}, $e^+e^- \to \tau^+\tau^-\gamma$ \cite{Acciarri} and $Z\to \tau^+\tau^-\gamma$ \cite{Ackerstaff,Grifols}. Furthermore,
an important point is the availability of high luminosity photon beams due to Bremsstrahlung as a byproduct in planned high energy colliders.
Also, very hard photons at high luminosity may be produced in Compton backscattering of laser light off high energy $e^+ e^-$ beams, as is
the case of the future CLIC.

In order to determine the bounds on the MM and EDM of the $\tau$-lepton, we calculate the total cross section of the reactions
$\gamma\gamma \to \tau^+\tau^-$ and $\gamma\gamma \to \tau^+\tau^-\gamma$. The most general parametrization for the electromagnetic
current between on-shell tau-lepton and the photon is given by \cite{Passera1,Grifols,Escribano,Giunti}

\begin{equation}
\Gamma^{\alpha}_\tau=eF_{1}(q^{2})\gamma^{\alpha}+\frac{ie}{2m_\tau}F_{2}(q^{2})\sigma^{\alpha\mu}q_{\mu}
+ \frac{e}{2m_\tau}F_3(q^2)\sigma^{\alpha\mu}q_\mu\gamma_5 +eF_4(q^2)\gamma_5(\gamma^\alpha - \frac{2q^\alpha m_\tau}{q^2}),
\end{equation}

\noindent where $e$ is the charge of the electron, $m_\tau$ is the mass of the tau-lepton, $\sigma^{\alpha\mu}=\frac{i}{2}[\gamma^\alpha, \gamma^{\mu}]$
represents the spin 1/2 angular momentum tensor and $q=p'-p$ is the momentum transfer. In the static (classical) limit the $q^2$-dependent
form factors $F_{1,2,3,4}(q^2)$ have familiar interpretations for $q^2=0$: $F_1(0)=Q_\tau$ is the electric charge; $F_2(0)=a_\tau$ its
anomalous MM and $F_3(0)=\frac{2m_\tau}{e} d_\tau$ with $d_\tau$ its EDM. $F_4(q^2)$ is the Anapole form factor.

In phenomenological and experimental searches most of the tau-lepton electromagnetic vertices search involve off-shell tau-leptons. Indeed
in these studies one of the tau-lepton is off-shell and measured quantity is not directly $a_\tau$ and $d_\tau$. For this reason deviations
of the tau-lepton dipole moments from the SM values are examined in an effective lagrangian approach. It is usually common to study new
physics in a model independent way through the effective lagrangian approach. This approach is defined by high-dimensional
operators which lead to anomalous $\tau^+ \tau^- \gamma$ coupling. In this study, we will apply the dimension-six effective operators that
contribute to the electric and magnetic dipole moments of the tau-lepton at the tree level given by Ref. \cite{1,eff1,eff3}:

\begin{eqnarray}
L_{eff}=\frac{1}{\Lambda^{2}} \Bigl[C_{LW}^{33} Q_{LW}^{33}+C_{LB}^{33} Q_{LB}^{33}\Bigr],
\end{eqnarray}

\noindent where

\begin{eqnarray}
Q_{LW}^{33}=\bigl(\bar{\ell_{\tau}}\sigma^{\mu\nu}\tau_{R}\bigr)\sigma^{I}\varphi W_{\mu\nu}^{I},
\end{eqnarray}

\begin{eqnarray}
Q_{LB}^{33}=\bigl(\bar{\ell_{\tau}}\sigma^{\mu\nu}\tau_{R}\bigr)\varphi B_{\mu\nu}.
\end{eqnarray}

Here, $\varphi$ and $\ell_{\tau}$ are the Higgs and the left-handed $SU(2)$ doublets, $\sigma^{I}$ are the Pauli
matrices and $W_{\mu\nu}^{I}$ and $B_{\mu\nu} $ are the gauge field strength tensors.

After the electroweak symmetry breaking from the effective lagrangian in Eq. (5), the CP even
$\kappa$ and CP odd $\tilde{\kappa}$ parameters are obtained

\begin{eqnarray}
\kappa=\frac{2 m_{\tau}}{e} \frac{\sqrt{2}\upsilon}{\Lambda^{2}} Re\Bigl[\cos\theta _{W} C_{LB}^{33}- \sin\theta _{W} C_{LW}^{33}\Bigr],
\end{eqnarray}

\begin{eqnarray}
\tilde{\kappa}=\frac{2 m_{\tau}}{e} \frac{\sqrt{2}\upsilon}{\Lambda^{2}} Im\Bigl[\cos\theta _{W} C_{LB}^{33}- \sin\theta _{W} C_{LW}^{33}\Bigr],
\end{eqnarray}

\noindent where $\upsilon=246$ GeV and $\sin\theta _{W}$ is the weak mixing angle.

These parameters are related to contribution of the anomalous magnetic and electric dipole moments of the tau-lepton through the following relations:

\begin{eqnarray}
\kappa=\tilde{a}_{\tau},
\end{eqnarray}

\begin{eqnarray}
\tilde{\kappa}=\frac{2m_{\tau}}{e}\tilde{d}_{\tau}.
\end{eqnarray}

\subsection{ $\gamma\gamma \to \tau^+\tau^-$ cross section}

All signal cross sections in this paper are computed using the package CALCHEP 3.6.30 \cite{Calhep}, which can computate the
Feynman diagrams, integrate over multiparticle phase space and event simulation. In addition, for our study we consider the
following basic acceptance cuts for $\tau^+\tau^-$ events at the CLIC:

\begin{eqnarray}
\begin{array}{c}
p^{\tau, \bar\tau}_t > 20\hspace{0.8mm}GeV,\\
|\eta^{\tau, \bar\tau}|< 2.5,\\
\Delta R(\tau, \bar\tau ) >0.4,\\
\end{array}
\end{eqnarray}

\noindent we apply these cuts to reduce the background and to optimize the signal sensitivity. In Eq. (12), $p^{\tau, \bar\tau}_t$
is the transverse momentum of the final state particles, $\eta^{\tau, \bar\tau}$ is the pseudorapidity which reduces the contamination
from other particles misidentified as tau and $\Delta R(\tau, \bar\tau)$ is the separation of the final state particles.

The tau-lepton was discovered by Martin Lewis Perl in 1975 \cite{Perl,Perl1}. It was discovered in the Stanford Positron and Electron Accelerator
Ring (SPEAR) of SLAC with the MARK I detector. With these tools, Perl and his team managed to distinguish leptons, hadrons and photons fairly accurately.
The tau-lepton was discovered from certain anomalies detected in the disintegration of the particles.
The observed event was as follows

\[
e^+e^- \to \tau^+\tau^- \to e^{\pm}+ \mu^{\pm}+\nu_e +\bar\nu_\mu + \nu_\tau +\bar\nu_\tau.
\]

When making the energy balance between the initial and final states, it was observed that the final energy was lower. At no time did the muons,
hadrons and photons sum the energy necessary to equal the initial state. Then it was thought that the energy that made the electron and the
positron collider created a pair of  new particles very massive, which soon decay into the other observed particles. This theory was difficult
to verify because the energy needed to produce the tau-antitau pair was similar to that required to create a pair of mesons. Subsequent experiments
carried out in DESY and in SLAC confirmed the existence of the tau lepton and provided more precise values for its mass and spin.

The tau is the only lepton that has the mass necessary to disintegrate, most of the time in hadrons. $17.8\%$ of the time the tau decays into an
electron and into two neutrinos; in another $17.4\%$ of the time, it decays in a muon and in two neutrinos. In the remaining $64.8\%$ of the occasions,
it decays in the form of hadrons and a neutrino. In Table II the main $\tau$-decay branching ratios are shown.

\begin{table}[!ht]
\caption{The main $\tau$-decay branching ratios \cite{Data2016,Bagliesi}.}
\begin{center}
\begin{tabular}{|c|| c|}
\hline
{\bf Channel}                                                               &     {\bf Branching Ratios $(\%)$}  \\
\hline\hline
$\tau \to e^-\nu_e\nu_\tau$                                                  &    $17.8 $   \\
\hline
$\tau \to \mu\nu_\mu \nu_\tau$                                               &    $17.4 $   \\
\hline
$\tau \to \pi^{\pm} \nu_\tau$                                                &    $11.1 $  \\
\hline
$\tau \to  \pi^0\pi^{\pm} \nu_\tau$                                          &    $25.4 $  \\
\hline
$\tau \to  \pi^0 \pi^0\pi^{\pm} \nu_\tau$                                    &    $9.19 $  \\
\hline
$\tau \to  \pi^0\pi^0 \pi^0\pi^{\pm} \nu_\tau$                               &    $1.08 $  \\
\hline
$\tau \to  \pi^{\pm}\pi^{\pm}\pi^{\pm} \nu_\tau$                             &    $8.98 $  \\
\hline
$\tau \to  \pi^0\pi^{\pm}\pi^{\pm}\pi^{\pm} \nu_\tau$                        &    $4.30 $  \\
\hline
$\tau \to  \pi^0\pi^0\pi^{\pm}\pi^{\pm}\pi^{\pm} \nu_\tau$                   &    $0.50 $  \\
\hline
$\tau \to  \pi^0\pi^0\pi^0\pi^{\pm}\pi^{\pm}\pi^{\pm} \nu_\tau$              &    $0.11 $  \\

\hline
$\tau \to  K^{\pm}X \nu_\tau$                                                &    $3.74 $  \\
\hline
$\tau \to  (\pi^0)\pi^{\pm}\pi^{\pm}\pi^{\pm}\pi^{\pm}\pi^{\pm} \nu_\tau$    &    $0.10 $  \\
\hline
Others                                                                       &    $0.03 $  \\
\hline
\end{tabular}
\end{center}
\end{table}

Since its discovery in 1975, the lepton-tau has been important to check different aspects of the SM. In particular,
since the tau is the charged lepton of the third generation, the verification of its properties can give light on the problems related to
the replication of generations, which is one of the open problems of the SM. Furthermore, it is the heavier lepton which makes
it especially more sensitive to new physics, since its coupling to the dynamics responsible for the generation of masses, whatever it is,
is more intense. In addition, it is the only lepton heavy enough to disintegrate into hadrons, which makes this particle a
particularly suitable system for studying Quantum Chromodynamics (QCD) at low energies.

The square matrix elements for the process $\gamma\gamma \to \tau^+\tau^-$ as a function of the Mandelstam invariants $\hat s$, $\hat t$
and $\hat u$ are given by:

\begin{eqnarray}
|M_{1}|^{2}&=&\frac{16\pi^{2}Q_{\tau}^2\alpha^{2}_e}{2m_{\tau}^{4}(\hat{t}-m_{\tau}^{2})^{2}}\biggl[48 \kappa (m_{\tau}^{2}-\hat{t})
(m_{\tau}^{2}+\hat{s}-\hat{t})m_{\tau}^{4}-16(3m_{\tau}^{4}-m_{\tau}^{2}\hat{s}+\hat{t}(\hat{s}+\hat{t})) m_{\tau}^{4}\nonumber\\
&+&2(m_{\tau}^{2}-\hat{t})( \kappa^{2}( 17m_{\tau}^{4}+(22\hat{s}-26\hat{t})m_{\tau}^{2} +\hat{t}(9\hat{t}-4\hat{s}))  \nonumber\\
&+&\tilde{\kappa}^{2}(17m_{\tau}^{2}+4\hat{s}-9\hat{t})(m_{\tau}^{2}-\hat{t}))m_{\tau}^{2}+12\kappa(\kappa^{2}
+\tilde{\kappa}^{2})\hat{s}(m_{\tau}^{3}-m_{\tau}\hat{t})^{2}\nonumber\\
&-&(\kappa^{2}+ \tilde{\kappa}^{2})^{2}(m_{\tau}^{2}-\hat{t})^{3}(m_{\tau}^{2}-\hat{s}-\hat{t})\biggr],
\end{eqnarray}

\begin{eqnarray}
|M_{2}|^{2}&=&\frac{-16\pi^{2}Q_{\tau}^2\alpha^{2}_e}{2m_{\tau}^{4}(\hat{u}-m_{\tau}^{2})^{2}}\biggl[48 \kappa (m_{\tau}^{4}+(\hat{s}-2\hat{t})m_{\tau}^{2}+\hat{t}(\hat{s}+\hat{t}))m_{\tau}^{4}\nonumber\\
&+&16(7m_{\tau}^{4}-(3\hat{s}+4\hat{t})m_{\tau}^{2}+\hat{t}(\hat{s}+\hat{t})) m_{\tau}^{4}\nonumber\\
&+&2(m_{\tau}^{2}-\hat{t})( \kappa^{2}(m_{\tau}^{4}+(17\hat{s}-10\hat{t})m_{\tau}^{2}+9\hat{t}(\hat{s}+\hat{t})) \nonumber\\
&+& \tilde{\kappa}^{2}(m_{\tau}^{2}-9\hat{t})(m_{\tau}^{2}-\hat{t}-\hat{s}))m_{\tau}^{2}\nonumber\\
&+&( \kappa^{2}+ \tilde{\kappa}^{2})^{2}(m_{\tau}^{2}-\hat{t})^{3}(m_{\tau}^{2}-\hat{s}-\hat{t})\biggr],
\end{eqnarray}

\begin{eqnarray}
M_{1}^{\dag}M_{2}+M_{2}^{\dag}M_{1}&=&\frac{16\pi^{2}Q_{\tau}^2\alpha^{2}_e}{m_{\tau}^{2}(\hat{t}-m_{\tau}^{2})(\hat{u}-m_{\tau}^{2})} \nonumber \\
&\times &\biggl[-16(4m_{\tau}^{6}-m_{\tau}^{4}\hat{s})+8 \kappa m_{\tau}^{2}(6m_{\tau}^{4}-6m_{\tau}^{2}(\hat{s}+2\hat{t})-\hat{s})^{2} \nonumber \\ &+&6\hat{t})^{2}+6\hat{s}\hat{t})+( \kappa^{2}(16m_{\tau}^{6}-m_{\tau}^{4}(15\hat{s}+32\hat{t})+m_{\tau}^{2}(15\hat{s})^{2} \nonumber \\
&+&14\hat{t}\hat{s}+16\hat{t})^{2})+\hat{s}\hat{t}(\hat{s}+\hat{t}))+ \tilde{\kappa}^{2}(16m_{\tau}^{6}-m_{\tau}^{4}(15\hat{s}+32\hat{t})  \nonumber\\
&+&m_{\tau}^{2}(5\hat{s})^{2}+14\hat{t}\hat{s}+16\hat{t})^{2})+\hat{s}\hat{t}(\hat{s}+\hat{t})))-4 \kappa \hat{s}( \kappa^{2}+ \tilde{\kappa}^{2})\nonumber\\
&\times& (m_{\tau}^{4}+m_{\tau}^{2}(\hat{s}-2\hat{t})+\hat{t}(\hat{s}+\hat{t}))-4 \tilde{\kappa} (\kappa^{2}+ \tilde{\kappa}^{2})(2m_{\tau}^{2}
-\hat{s}-2\hat{t}) \nonumber\\
&\times& \epsilon_{\alpha \beta \gamma \delta} p_{1}^{\alpha}p_{2}^{\beta}p_{3}^{\gamma}p_{4}^{\delta}-2\hat{s}(\kappa^{2}+ \tilde{\kappa}^{2})^{2}
(m_{\tau}^{4}-2\hat{t}m_{\tau}^{2}+\hat{t}(\hat{s}+\hat{t}))\biggr],
\end{eqnarray}

\noindent where $\hat s=(p_1 + p_2)^2=(p_3 + p_4)^2$, $\hat t=(p_1 - p_3)^2=(p_4 - p_2)^2$, $\hat u=(p_3 - p_2)^2=(p_1 - p_4)^2$, and
$p_{1}$ and $p_{2}$ are the four-momenta of the incoming photons, $p_{3}$ and $p_{4}$ are the momenta of the outgoing tau-lepton,
$Q_{\tau}$ is the tau-lepton charge, $\alpha_e$ is the fine-structure constant and $m_\tau$ is the mass of the tau.

The most promising mechanism to generate energetic photon beams in a linear collider is Compton backscattering. Compton backscattered
photons interact with each other and generate the process $\gamma \gamma \rightarrow \tau^+ \tau^-$. The spectrum of Compton backscattered
photons is given by

 \begin{eqnarray}
 f_{\gamma}(y)=\frac{1}{g(\zeta)}[1-y+\frac{1}{1-y}-
 \frac{4y}{\zeta(1-y)}+\frac{4y^{2}}{\zeta^{2}(1-y)^{2}}] ,
 \end{eqnarray}

 where

 \begin{eqnarray}
 g(\zeta)=(1-\frac{4}{\zeta}-\frac{8}{\zeta^2})\log{(\zeta+1)}+
 \frac{1}{2}+\frac{8}{\zeta}-\frac{1}{2(\zeta+1)^2} ,
 \end{eqnarray}

 with

 \begin{eqnarray}
 y=\frac{E_{\gamma}}{E_{e}} , \;\;\;\; \zeta=\frac{4E_{0}E_{e}}{M_{e}^2}
 ,\;\;\;\; y_{max}=\frac{\zeta}{1+\zeta}.
 \end{eqnarray}

Here, $E_{0}$ and $E_{e}$ are  energy of the incoming laser photon and initial energy of the electron beam before
Compton backscattering and $E_{\gamma}$ is the energy of the backscattered photon. The maximum value of $y$ reaches
0.83 when $\zeta=4.8$.

The total cross section is given by,

\begin{eqnarray}
\sigma=\int f_{\gamma}(x)f_{\gamma}(x)d\hat{\sigma}dE_{1}dE_{2}.
\end{eqnarray}

Next, we present the total cross section as a polynomial in powers of $\kappa ( \tilde{\kappa})$. This provides more precise and convenient
information for the study of the process $\gamma \gamma \rightarrow \tau^+ \tau^-$. We consider the following cases:\\

$\bullet$ For $\sqrt{s}=380\hspace{0.8mm} GeV$.

\begin{eqnarray}
\sigma(\kappa)&=&\big[(9.73\times 10^6) \kappa^4 + (8.18\times 10^4) \kappa^3 + (8.13\times 10^4) \kappa^2 + (1.11\times 10^2) \kappa + 38.75\bigr](pb),   \nonumber \\
\sigma(\tilde{\kappa})&=&\bigl[(9.73\times 10^6) \tilde{\kappa}^4 + (8.26\times 10^4) \tilde{\kappa}^2 + 38.75\bigr](pb).
\end{eqnarray}

$\bullet$ For $\sqrt{s}=1500\hspace{0.8mm} GeV$.

\begin{eqnarray}
\sigma(\kappa)&=&\bigl[(1.54 \times 10^8) \kappa^4 + (8.42 \times 10^8) \kappa^3 + (8.80\times 10^4) \kappa^2 + (17.5)\kappa + 6\bigr](pb),  \nonumber  \\\
\sigma(\tilde{\kappa})&=&\bigl[(1.54 \times 10^{8}) \tilde{\kappa}^4 + (8.81 \times 10^4) \tilde{\kappa}^2 + 6\bigr](pb).
\end{eqnarray}

$\bullet$ For $\sqrt{s}=3000\hspace{0.8mm} GeV$.

\begin{eqnarray}
\sigma(\kappa)&=&\bigl[(6.17 \times 10^{8}) \kappa^4 + (9.13\times 10^4) \kappa^3 + (8.72 \times 10^4) \kappa^2 - (1.21)\kappa + 1.97\bigr](pb),   \nonumber  \\
\sigma(\tilde{\kappa})&=&\bigl[(6.17 \times 10^{8}) \tilde{\kappa}^4 + (8.83\times 10^4) \tilde{\kappa}^2 + 1.97\bigr](pb).
\end{eqnarray}

From Eqs. (20)-(22), the linear, quadratic and cubic terms in $\kappa (\tilde{\kappa})$ arise from the interference between SM and anomalous amplitudes,
whereas the quartic terms are purely anomalous. The independent term of $\kappa (\tilde{\kappa})$ correspond to the cross section
at $\kappa=\tilde{\kappa}=0$ and represents the contribution of the cross section of the SM.

\subsection{Bounds on the $\tilde{a}_\tau$ and $\tilde{d}_\tau$ through $\gamma \gamma \rightarrow \tau^+ \tau^-$ at the CLIC}

For our numerical analysis of the total cross section $\sigma_{NP}(\gamma \gamma \rightarrow \tau^+ \tau^-)=\sigma_{NP}(\sqrt{s},\hspace{0.8mm} \kappa,\hspace{0.8mm}\tilde{\kappa})$, as well as of the electromagnetic dipole moments of the tau-lepton, where the free parameters are the
center-of-mass energy $\sqrt{s}$, the integrated luminosity ${\cal L}$ of the CLIC and the factors $\kappa$ and $\tilde{\kappa}$, we
also consider the acceptance cuts given in Eq. (12). In addition, we take into account the systematic uncertainties for the collider.
For this purpose, we use the usual formula for the $\chi^2$ function \cite{Koksal1,Ozguven,Billur,Sahin1}:

\begin{equation}
\chi^2=\biggl(\frac{\sigma_{SM}-\sigma_{NP}(\sqrt{s}, \kappa, \tilde{\kappa})}{\sigma_{SM}\delta}\biggr)^2,
\end{equation}

\noindent where $\sigma_{NP}(\sqrt{s}, \kappa, \tilde{\kappa})$ is the total cross section including contributions from the SM
and new physics, $\delta=\sqrt{(\delta_{st})^2+(\delta_{sys})^2}$, $\delta_{st}=\frac{1}{\sqrt{N_{SM}}}$
is the statistical error, $\delta_{sys}$ is the systematic error and $N_{SM}$ is the number of signal expected
events $N_{SM}={\cal L}_{int}\times BR \times \sigma_{SM}$ where ${\cal L}_{int}$ is the integrated CLIC luminosity.
Furthermore, as the tau-lepton decays roughly $35\%$ of the time leptonically and $65\%$ of the time to one or more
hadrons (see Table II), then for the signal we consider one of the tau-leptons decays leptonically and the other
hadronically. Therefore, for our study we assume that the Branching ratio of the two-tau in the final state to be BR=0.46.

Systematic uncertainties may occur when tau-lepton is identified due to some of the reasons described below:
Although, we do not have any CLIC reports \cite{6,7,8} to know exactly what the systematic uncertainties are for our processes which
are investigated, we will make some approaches about the systematic uncertainties. The DELPHI Collaboration examined the anomalous
magnetic and electric dipole moments of the tau-lepton through the process $e^{+}e^{-} \rightarrow
e^{+}e^{-}\tau^{+}\tau^{-}$ in the years $1997-2000$ at collision energy $\sqrt{s}$ between $183$ and $208$ GeV \cite{Abdallah}. Relative
systematic errors on cross section of the process $e^{+}e^{-} \rightarrow e^{+}e^{-}\tau^{+}\tau^{-}$ are given in Table \ref{tabex}.
Also, the process $e^{+}e^{-} \rightarrow e^{+}e^{-}\tau^{+}\tau^{-}$ was studied with the L3 detector for center-of-mass energies
$161$ GeV$\leqslant \sqrt{s} \leqslant 209$ GeV at LEP \cite{Acciarri}. The anomalous magnetic and electric dipole moments of the tau-lepton
via the process $p p \rightarrow p p\tau^{+}\tau^{-}$ with $2\%$ of the total systematic uncertainties at the LHC was
investigated phenomenologically in Ref \cite{Atag}. Work in this regard is done by ATLAS and CMS groups \cite{atlastau1,atlastau2,Bagliesi}.
Tau tagging efficiencies also studied for the International Large Detector (ILD) \cite{ild}, a proposed detector concept for the
International Linear Collider (ILC). Due to these difficulties, tau identification efficiencies are always calculated
for specific process, luminosity, and kinematic parameters. These studies are currently being carried out by various groups for selected
productions. For a realistic efficiency, we need a detailed study for our specific process and kinematic parameters. For all these reasons,
in this work, kinematic cuts contain some general values chosen by detectors for lepton identification. Hence, in this paper, tau-lepton
identification efficiency is considered within systematic errors. Taking into consideration the previous studies, $3\%$ and $5\%$ of total
systematic uncertainties were taken in this study. It can be assumed that this accelerator will be built in the coming years and the systematic
uncertainties will be lower when considering the development of future detector technology.

\begin{table}
\caption{Systematic errors given by the DELPHI Collaboration \cite{Abdallah}.
\label{tabex}}
\begin{ruledtabular}
\begin{tabular}{ccccc}
 & $1997$& $1998$& $1999$& $2000$ \\
\hline
Trigger efficiency& $7.0$& $2.7$& $3.6$& $4.5$ \\
Selection efficiency& $5.1$& $3.2$& $3.0$& $3.0$  \\
Background& $1.7$& $0.9$& $0.9$& $0.9$  \\
Luminosity& $0.6$& $0.6$& $0.6$& $0.6$ \\
Total& $8.9$& $4.3$& $4.7$& $5.4$  \\
\end{tabular}
\end{ruledtabular}
\end{table}

With all these elements that we taken into consideration, we made and presented a set of figures, as well as tables which
illustrate our results.

The total cross sections $\sigma_{\gamma \gamma \rightarrow \tau^+ \tau^-}(\sqrt{s}, \kappa, \tilde{\kappa})$ are presented as a function of
the anomalous couplings $\kappa$ in Fig. 3 and $\tilde{\kappa}$ in Fig. 4 for the center-of-mass energies of $\sqrt{s}= 380, 1500, 3000\hspace{0.8mm}GeV$,
respectively. The total cross section clearly shows a strong dependence with respect to the anomalous parameters $\kappa$, $\tilde{\kappa}$
and with the center-of-mass energy of the collider $\sqrt{s}$. Additionally, the $\sigma_{\gamma \gamma \rightarrow \tau^+ \tau^-}(\sqrt{s},
\kappa, \tilde{\kappa})$ as function of $\kappa$ and $\tilde{\kappa}$ are shown in Figs. 5-7. In these figures, the surfaces are increased
for the lower and upper limits of the parameters $\kappa$ and $\tilde{\kappa}$, showing a strong dependence with respect to these parameters.

$95\%$  C. L. allowed regions in the plane $(\kappa-\tilde{\kappa})$ for the process $\gamma \gamma \rightarrow \tau^+ \tau^-$ for the first,
second and third stage of operation of the CLIC, where a fixed center-of-mass energies of $\sqrt{s}= 380, 1500, 3000\hspace{0.8mm}GeV$ is
assumed with luminosities of ${\cal L}=10, 100, 500\hspace{0.8mm}fb^{-1}$, ${\cal L}=100, 500, 1500\hspace{0.8mm}fb^{-1}$
and ${\cal L}=100, 500, 3000\hspace{0.8mm}fb^{-1}$, respectively, and considering systematic uncertainties of $\delta_{sys}=0\%, 3\%, 5\hspace{0.8mm}\%$
\cite{Abdallah,Achard}, they are displayed in Figs. 8-10. For a complementary description on the uncertainties we suggest the reader consult
Refs. \cite{Chatrchyan,Atag}.

The achievable precision in the determination of the anomalous magnetic moment $\tilde{a}_\tau$ and the electric dipole moment $\tilde{d}_\tau$
are summarized in Figs. 11-14 and Tables IV-VI and are compared with experimental results of earlier studios for a linear collider as published
by the BELLE, DELPHI, L3 and OPAL Collaboration \cite{Inami,Abdallah,Acciarri,Ackerstaff,Albrecht}. Our results show that the two-photon process
$\gamma \gamma \rightarrow \tau^+ \tau^-$ at the CLIC improve the sensitivity bounds on anomalous electromagnetic dipole moments of tau-lepton
with respect to the existing experimental bounds (see Table I) by two orders of magnitude. Our best bounds obtained on $\tilde{a}_\tau$ and
$\tilde{d}_\tau$ are $-0.00015\leq \tilde{a}_\tau \leq 0.00017$ and $|\tilde{d}_\tau(ecm)|=9.040\times 10^{-19}$, respectively, as shown in
Tables IV-VI.

\begin{table}[!ht]
\caption{Sensitivity on the $\tilde{a}_\tau$ magnetic moment and the $\tilde{d}_\tau$ electric dipole moment for $\sqrt{s}=380\hspace{0.8mm}GeV$
and ${\cal L}=10, 50, 100, 300, 500\hspace{0.8mm}fb^{-1}$ at $95\%$ C.L. through the process $\gamma\gamma \to \tau \bar\tau$.}
\begin{center}
 \begin{tabular}{ccccc}
\hline\hline
\multicolumn{4}{c}{ $\sqrt{s}=380\hspace{0.8mm}GeV$, \hspace{5mm}  $95\%$ C.L.}\\
 \hline
 \cline{1-4}  ${\cal L}\hspace{0.8mm}(fb^{-1})$  & \hspace{1.5cm} $\delta_{sys}$ & \hspace{1.5cm}    $\tilde{a}_\tau$   & \hspace{1.7cm}
 $|\tilde{d}_\tau(e cm)|$ \\
\hline
10  &\hspace{1.2cm}  $0\%$   &\hspace{1.2cm} [-0.00232; 0.00095]   & \hspace{1.5cm}      $1.071\times 10^{-17}$   \\
10  &\hspace{1.2cm}  $3\%$   &\hspace{1.2cm} [-0.00603; 0.00464]    & \hspace{1.5cm}     $2.999\times 10^{-17}$   \\
10  &\hspace{1.2cm}  $5\%$   &\hspace{1.2cm} [-0.00756; 0.00616]    & \hspace{1.5cm}     $3.826\times 10^{-17}$   \\
\hline
50   &\hspace{1.2cm} $0\%$   &\hspace{1.2cm} [-0.00189; 0.00052]    & \hspace{1.5cm}     $8.813\times 10^{-18}$  \\
50  &\hspace{1.2cm}  $3\%$   &\hspace{1.2cm} [-0.00603; 0.00464]    & \hspace{1.5cm}     $2.998\times 10^{-17}$  \\
50  &\hspace{1.2cm}  $5\%$   &\hspace{1.2cm} [-0.00756; 0.00616]    & \hspace{1.5cm}     $3.825\times 10^{-17}$  \\
\hline
100   &\hspace{1.2cm} $0\%$  &\hspace{1.2cm} [-0.00176; 0.00039]    & \hspace{1.5cm}     $8.298\times 10^{-18}$  \\
100  &\hspace{1.2cm}  $3\%$  &\hspace{1.2cm} [-0.00603; 0.00464]    & \hspace{1.5cm}     $2.998\times 10^{-17}$  \\
100  &\hspace{1.2cm}  $5\%$  &\hspace{1.2cm} [-0.00756; 0.00616]    & \hspace{1.5cm}     $3.825\times 10^{-17}$  \\
\hline
300   &\hspace{1.2cm} $0\%$   &\hspace{1.2cm} [-0.00161; 0.00025]    & \hspace{1.5cm}    $7.737\times 10^{-18}$  \\
300  &\hspace{1.2cm}  $3\%$   &\hspace{1.2cm} [-0.00603; 0.00464]    & \hspace{1.5cm}    $2.998\times 10^{-17}$  \\
300  &\hspace{1.2cm}  $5\%$   &\hspace{1.2cm} [-0.00756; 0.00616]    & \hspace{1.5cm}    $3.825\times 10^{-17}$  \\
\hline
500   &\hspace{1.2cm} $0\%$  &\hspace{1.2cm} [-0.00156; 0.00019]    & \hspace{1.5cm}     $7.556\times 10^{-18}$  \\
500  &\hspace{1.2cm}  $3\%$  &\hspace{1.2cm} [-0.00603; 0.00464]    & \hspace{1.5cm}     $2.997\times 10^{-17}$  \\
500  &\hspace{1.2cm}  $5\%$  &\hspace{1.2cm} [-0.00756; 0.00616]    & \hspace{1.5cm}     $3.825\times 10^{-17}$  \\
\hline\hline
\end{tabular}
\end{center}
\end{table}

\begin{table}[!ht]
\caption{Sensitivity on the $\tilde{a}_\tau$ magnetic moment and the $\tilde{d}_\tau$ electric dipole moment for $\sqrt{s}=1.5\hspace{0.8mm}TeV$
and ${\cal L}=100, 300, 500, 1000, 1500\hspace{0.8mm}fb^{-1}$ at $95\%$ C.L. through the process $\gamma\gamma \to \tau \bar\tau$.}
\begin{center}
 \begin{tabular}{ccccc}
\hline\hline
\multicolumn{4}{c}{ $\sqrt{s}=1.5\hspace{0.8mm}TeV$, \hspace{5mm}  $95\%$ C.L.}\\
 \hline
 \cline{1-4}  ${\cal L}\hspace{0.8mm}(fb^{-1})$  & \hspace{1.5cm} $\delta_{sys}$ & \hspace{1.5cm}     $\tilde{a}_\tau$   & \hspace{1.7cm}
 $|\tilde{d}_\tau(e cm)|$ \\
\hline
100  &\hspace{1.2cm}  $0\%$   &\hspace{1.2cm} [-0.00061; 0.00041]   & \hspace{1.5cm}     $2.796\times 10^{-18}$   \\
100  &\hspace{1.2cm}  $3\%$   &\hspace{1.2cm} [-0.00209; 0.00189]    & \hspace{1.5cm}    $1.109\times 10^{-17}$   \\
100  &\hspace{1.2cm}  $5\%$   &\hspace{1.2cm} [-0.00267; 0.00247]    & \hspace{1.5cm}    $1.428\times 10^{-17}$   \\
\hline
300   &\hspace{1.2cm} $0\%$   &\hspace{1.2cm} [-0.00049; 0.00029]    & \hspace{1.5cm}    $2.1226\times 10^{-18}$  \\
300  &\hspace{1.2cm}  $3\%$   &\hspace{1.2cm} [-0.00209; 0.00189]    & \hspace{1.5cm}    $1.108\times 10^{-17}$  \\
300  &\hspace{1.2cm}  $5\%$   &\hspace{1.2cm} [-0.00267; 0.00247]    & \hspace{1.5cm}    $1.428\times 10^{-17}$  \\
\hline
500   &\hspace{1.2cm} $0\%$  &\hspace{1.2cm} [-0.00044; 0.00025]    & \hspace{1.5cm}     $1.867\times 10^{-18}$  \\
500  &\hspace{1.2cm}  $3\%$  &\hspace{1.2cm} [-0.00209; 0.00189]    & \hspace{1.5cm}     $1.108\times 10^{-17}$  \\
500  &\hspace{1.2cm}  $5\%$  &\hspace{1.2cm} [-0.00267; 0.00247]    & \hspace{1.5cm}     $1.428\times 10^{-17}$  \\
\hline
1000   &\hspace{1.2cm} $0\%$   &\hspace{1.2cm} [-0.00039; 0.00019]    & \hspace{1.5cm}   $1.567\times 10^{-18}$  \\
1000  &\hspace{1.2cm}  $3\%$   &\hspace{1.2cm} [-0.00209; 0.00189]    & \hspace{1.5cm}   $1.108\times 10^{-17}$  \\
1000  &\hspace{1.2cm}  $5\%$   &\hspace{1.2cm} [-0.00267; 0.00247]    & \hspace{1.5cm}   $1.427\times 10^{-17}$  \\
\hline
1500   &\hspace{1.2cm} $0\%$  &\hspace{1.2cm} [-0.00037; 0.00017]    & \hspace{1.5cm}    $1.415\times 10^{-18}$  \\
1500  &\hspace{1.2cm}  $3\%$  &\hspace{1.2cm} [-0.00209; 0.00189]    & \hspace{1.5cm}    $1.108\times 10^{-17}$  \\
1500  &\hspace{1.2cm}  $5\%$ &\hspace{1.2cm}  [-0.00267; 0.00247]    & \hspace{1.5cm}    $1.427\times 10^{-17}$  \\
\hline\hline
\end{tabular}
\end{center}
\end{table}

\begin{table}[!ht]
\caption{Sensitivity on the $\tilde{a}_\tau$ magnetic moment and the $\tilde{d}_\tau$ electric dipole moment for $\sqrt{s}=3\hspace{0.8mm}TeV$
and ${\cal L}=100, 500, 1000, 2000, 3000\hspace{0.8mm}fb^{-1}$ at $95\%$ C.L. through the process $\gamma\gamma \to \tau \bar\tau$.}
\begin{center}
 \begin{tabular}{ccccc}
\hline\hline
\multicolumn{4}{c}{ $\sqrt{s}=3\hspace{0.8mm}TeV$, \hspace{5mm}  $95\%$ C.L.}\\
 \hline
 \cline{1-4}  ${\cal L}\hspace{0.8mm}(fb^{-1})$  & \hspace{1.5cm} $\delta_{sys}$ & \hspace{1.5cm} $\tilde{a}_\tau$   & \hspace{1.7cm}
 $|\tilde{d}_\tau(e cm)|$ \\
\hline
100  &\hspace{1.2cm}  $0\%$   &\hspace{1.2cm} [-0.00037; 0.00039]   & \hspace{1.5cm}     $2.116\times 10^{-18}$   \\
100  &\hspace{1.2cm}  $3\%$   &\hspace{1.2cm} [-0.00114; 0.00115]    & \hspace{1.5cm}    $6.342\times 10^{-18}$   \\
100  &\hspace{1.2cm}  $5\%$   &\hspace{1.2cm} [-0.00147; 0.00148]    & \hspace{1.5cm}    $8.153\times 10^{-18}$   \\
\hline
500   &\hspace{1.2cm} $0\%$   &\hspace{1.2cm} [-0.00025; 0.00026]    & \hspace{1.5cm}    $1.415\times 10^{-18}$  \\
500  &\hspace{1.2cm}  $3\%$   &\hspace{1.2cm} [-0.00114; 0.00115]    & \hspace{1.5cm}    $6.335\times 10^{-18}$  \\
500  &\hspace{1.2cm}  $5\%$  &\hspace{1.2cm}  [-0.00147; 0.00148]    & \hspace{1.5cm}    $8.153\times 10^{-18}$  \\
\hline
1000   &\hspace{1.2cm} $0\%$  &\hspace{1.2cm} [-0.00020; 0.00022]    & \hspace{1.5cm}    $1.190\times 10^{-18}$  \\
1000  &\hspace{1.2cm}  $3\%$  &\hspace{1.2cm} [-0.00114; 0.00115]    & \hspace{1.5cm}    $6.334\times 10^{-18}$  \\
1000  &\hspace{1.2cm}  $5\%$ &\hspace{1.2cm}  [-0.00147; 0.00148]    & \hspace{1.5cm}    $8.153\times 10^{-18}$  \\
\hline
2000   &\hspace{1.2cm} $0\%$   &\hspace{1.2cm} [-0.00017; 0.00018]    & \hspace{1.5cm}   $1.000\times 10^{-18}$  \\
2000  &\hspace{1.2cm}  $3\%$   &\hspace{1.2cm} [-0.00114; 0.00115]    & \hspace{1.5cm}   $6.334\times 10^{-18}$  \\
2000  &\hspace{1.2cm}  $5\%$  &\hspace{1.2cm}  [-0.00147; 0.00148]    & \hspace{1.5cm}   $8.153\times 10^{-18}$  \\
\hline
3000   &\hspace{1.2cm} $0\%$  &\hspace{1.2cm} [-0.00015; 0.00017]    & \hspace{1.5cm}    $9.040\times 10^{-19}$  \\
3000  &\hspace{1.2cm}  $3\%$  &\hspace{1.2cm} [-0.00114; 0.00115]    & \hspace{1.5cm}    $6.334\times 10^{-18}$  \\
3000  &\hspace{1.2cm}  $5\%$ &\hspace{1.2cm}  [-0.00147; 0.00148]    & \hspace{1.5cm}    $8.153\times 10^{-18}$  \\
\hline\hline
\end{tabular}
\end{center}
\end{table}

%\newpage

\subsection{$\gamma\gamma \to \tau^+\tau^-\gamma$ cross section}

Experimentally, the processes that involving single-photon in the final state $(\tau^+\tau^-\gamma)$  can potentially distinguish
from background associated with the process under consideration. Furthermore, the anomalous $\tau^+\tau^-\gamma$ coupling can be analyzed
through the process $e^+e^- \to \tau^+\tau^- $ at the linear colliders. This process receives contributions from both anomalous $\tau^+\tau^- \gamma$
and $ \tau^+\tau^- Z$  couplings. However, the processes $\gamma\gamma \to \tau^+\tau^-$ and $\gamma\gamma \to \tau^+\tau^-\gamma$ isolate
$\tau^+\tau^- \gamma$ coupling which provides the possibility to analyze the $\tau^+ \tau^-\gamma$ coupling separately from the $\tau^+ \tau^- Z$
coupling. In general, anomalous values of $\tilde{a}_\tau$ and $\tilde{d}_\tau$ tend to increase the cross section for the process
$\gamma\gamma \to \tau^+\tau^-\gamma$, especially for photons with high energy which are well isolated from the decay products of the taus \cite{Acciarri}.
Additionally, the single-photon in the final state has the advantage of being identifiable with high efficiency and purity.

They may also provide a clear signal in the detector for new physics, for new phenomena such as the dipole moments of fermions.
Also, the selection criteria used for the analysis allow the search for events having the characteristics of single-photon.

We now turn attention to the process $\gamma\gamma \to \tau^+\tau^-\gamma$ at future $e^+e^-$ collider. On the technical side,
for the calculation of the total cross section of $\gamma\gamma \to \tau^+\tau^-\gamma$, the analytical expression for
the amplitude square is quite lengthy so we do not present it here. Instead, we present numerical fit functions for the total cross
sections with respect to center-of-mass energy and in terms of the parameters $\kappa$ and $\tilde{\kappa}$.
Furthermore, in the case of the process $\gamma\gamma \to \tau^+\tau^-\gamma$, we apply the following kinematic
cuts to reduce the background and to maximize the signal sensitivity:

\begin{eqnarray}
\begin{array}{c}
p^\gamma_t > 20 \hspace{0.8mm}GeV, \hspace{5mm} |\eta^{\gamma}|< 2.5,\\
p^{\tau, \bar\tau}_t > 20\hspace{0.8mm}GeV,    \hspace{5mm} |\eta^{\tau, \bar\tau}|< 2.5,\\
\Delta R(\tau, \gamma) >0.4,\\
\Delta R(\tau, \bar\tau) >0.4,\\
\Delta R(\bar\tau, \gamma) >0.4,\\
\end{array}
\end{eqnarray}

\noindent for the photon transverse momentum $p^\gamma_t$, the photon pseudorapidity $\eta^\gamma$ which reduces the contamination
from other particles misidentified as photon, the tau transverse momentum $p^{\tau, \bar\tau}_t$ for the final state particles,
the tau pseudorapidity $\eta^\tau$ which reduces the contamination from other particles misidentified as tau and $\Delta R(\tau, \gamma)$,
$\Delta R(\tau, \bar\tau)$ and $\Delta R(\bar\tau, \gamma)$ are the separation of the final state particles. In conclusion, by using these
cuts given in Eq. (24) in our manuscript we have taken into account isolation criteria to optimize the signal to the particles of the
$\tau^+\tau^-\gamma$ final state. The cases considered are:\\

\newpage

$\bullet$ For $\sqrt{s}=380\hspace{0.8mm} GeV$.

\begin{eqnarray}
\sigma(\kappa)&=&\bigl[(3.22\times 10^7)\kappa^6 + (4.08 \times 10^4)\kappa^5 + (3.36 \times 10^5)\kappa^4 + (1.87 \times 10^3) \kappa^3   \nonumber\\
           &+& (1.24\times 10^3) \kappa^2 + (0.515) \kappa + 0.21\bigl](pb),    \\
\sigma(\tilde{\kappa})&=& \bigl[(3.22\times 10^7)\tilde{\kappa}^6 + (3.43 \times 10^5) \tilde{\kappa}^4 + (1.24 \times 10^3) \tilde{\kappa}^2
+ 0.21\bigl](pb).
\end{eqnarray}

$\bullet$ For $\sqrt{s}=1500\hspace{0.8mm} GeV$.

\begin{eqnarray}
\sigma(\kappa)&=&\bigl[(8.29\times 10^9)\kappa^6 + (8.65 \times 10^6)\kappa^5 + (1.15 \times 10^7)\kappa^4 + (3.90 \times 10^3) \kappa^3   \nonumber\\
           &+& (3.74\times 10^3) \kappa^2 + (0.31) \kappa + 0.116\bigl](pb),    \\
\sigma(\tilde{\kappa})&=& \bigl[(8.29\times 10^9)\tilde{\kappa}^6 + (1.15 \times 10^7) \tilde{\kappa}^4 + (3.74 \times 10^3) \tilde{\kappa}^2
+ 0.116\bigl](pb).
\end{eqnarray}

$\bullet$ For $\sqrt{s}=3000\hspace{0.8mm} GeV$.

\begin{eqnarray}
\sigma(\kappa)&=&\bigl[(1.33\times 10^{11})\kappa^6 + (4.63 \times 10^7)\kappa^5 + (5.58 \times 10^7)\kappa^4 + (1.46 \times 10^3) \kappa^3   \nonumber\\
           &+& (4.92\times 10^3) \kappa^2 + (0.47) \kappa + 0.052\bigl](pb),    \\
\sigma(\tilde{\kappa})&=&\bigl[ (1.33\times 10^{11})\tilde{\kappa}^6 + (5.58 \times 10^7) \tilde{\kappa}^4 + (4.92 \times 10^3) \tilde{\kappa}^2
+ 0.052\bigl](pb).
\end{eqnarray}

\noindent It is worth mentioning that in the equations for the total cross section (25)-(30), the coefficients of $\kappa (\tilde{\kappa})$ given the
anomalous contribution, while the independent terms of $\kappa (\tilde{\kappa})$ correspond to the cross section at $\kappa=\tilde{\kappa}=0$ and represents the SM
cross section magnitude.

\subsection{Bounds on $\tilde{a}_\tau$ and $\tilde{d}_\tau$ through $\gamma\gamma \to \tau^+\tau^-\gamma$ at the CLIC}

Next, we focus our attention on the numerical calculation for the total cross section and for the electromagnetic dipole
moments of the tau-lepton. To carry out our objective, we start with Eqs. (25)-(30) and adopting the collider parameters of
$\sqrt{s}= 380, 1500, 3000 \hspace{0.8mm}GeV$ for the center-of-mass energy and ${\cal L}=10, 50, 100, 300, 500, 1000, 1500,
2000, 3000\hspace{0.8mm}fb^{-1}$ for the integrated luminosity of data. In addition, we apply kinematic cuts given by Eq. (24)
to optimize the signal and to reduce the background, and an important part of our study is the inclusion of systematic
uncertainties of $\delta_{sys}=0\%, 3\%, 5\%$. Our calculations yield the following results.

Figs. 15 and 16 present the results for the total cross section for the $\gamma\gamma \to \tau^+\tau^-\gamma$ scenario,
where the total cross sections for $\sigma(\gamma\gamma \to \tau^+\tau^-\gamma)$ vs $\kappa(\tilde{\kappa})$ are shown for
$\sqrt{s}= 380, 1500, 3000 \hspace{0.8mm}GeV$,
respectively. In both cases, the dependence of the total cross section $\sigma(\sqrt{s}, \kappa, \tilde{\kappa})$ presents a clear dependence
with respect to $\sqrt{s}$, as well as with the anomalous parameters $\kappa$ and $\tilde{\kappa}$. For case of comparison the total cross section
for $\gamma\gamma \to \tau^+\tau^-$ is major by a factor of 10 compared to the $\gamma\gamma \to \tau^+\tau^-\gamma$ scenario.
To visualize the effects of $\kappa$ and $\tilde{\kappa}$ on the total cross section $\sigma_{\gamma\gamma \to \tau^+\tau^-\gamma}(\sqrt{s}, \kappa, \tilde{\kappa})$ we display Figs. 17-19.

The contours plots on $\tilde{\kappa}$ and $\tilde{\kappa}$ at the $95\%$ C.L., are obtained using Eqs. (25)-(30), these are shown
in Figs. 20-22 for integrate luminosities of ${\cal L}=10, 100, 500, 1500, 3000\hspace{0.8mm}fb^{-1}$ with
$\sqrt{s}=380, 1500, 3000\hspace{0.8mm} GeV$, respectively. The improvement in the sensitivity on $\tilde{\kappa}$ and $\tilde{\kappa}$
is obtained using high energy and high luminosity as is immediately apparent from these figures.

The estimated sensitivity of the CLIC for $\tilde{a}_\tau (\tilde{d}_\tau)$ at $95\%$ C.L., as well as for different center-of-mass energies,
luminosities and systematic errors is illustrated in Figs. 23-26. The sensitivity is 1-2 orders of magnitude better for all couplings than that
expected at the the current colliders (see Table I) and the potential to disentangle $\tilde{a}_\tau$ and $\tilde{d}_\tau$ improves at larger
$\sqrt{s}$ and ${\cal L}$.

To conclude with our set of results, we present, through Tables VII-IX, the bounds corresponding to the magnetic and electric dipole
moments of the $\tau$-lepton, via the $\gamma\gamma \to \tau^+\tau^-\gamma$ mode. The center-of-mass energies and luminosities assumed
are $\sqrt{s}=380, 1500, 3000\hspace{0.8mm}$ and ${\cal L}=10, 50, 100, 300, 500, 1000, 1500, 2000, 3000\hspace{0.8mm}fb^{-1}$,
respectively, and only one anomalous coupling $\tilde{a}_\tau (\tilde{d}_\tau)$ was varied at a time to bound its value. In the present study,
the sensitivities are based on the systematic errors of $\delta_{sys}=0\%, 3\%, 5\%$ and our best bounds are:
$-0.00033\leq \tilde{a}_\tau \leq 0.00023$ and $|\tilde{d}_\tau(ecm)|=1.546\times 10^{-18}$, respectively. These bounds are weaker by an order of
magnitude than those corresponding to the $\gamma\gamma \to \tau^+\tau^-$ mode (see Subsection B, Tables IV-VI). But stronger than
those reported by BELLE, DELPHI, L3 and OPAL Collaborations (see Table 1).

\begin{table}[!ht]
\caption{Sensitivity on the $\tilde{a}_\tau$ magnetic moment and the $\tilde{d}_\tau$ electric dipole moment for $\sqrt{s}=380\hspace{0.8mm}GeV$
and ${\cal L}=10, 50, 100, 300, 500\hspace{0.8mm}fb^{-1}$ at $95\%$ C.L. through the process $\gamma\gamma \to \tau \bar\tau\gamma$.}
\begin{center}
 \begin{tabular}{ccccc}
\hline\hline
\multicolumn{4}{c}{$\sqrt{s}=380\hspace{0.8mm}GeV$, \hspace{5mm}  $95\%$ C.L.}\\
 \hline
 \cline{1-4}  ${\cal L}\hspace{0.8mm}(fb^{-1})$  & \hspace{1.5cm} $\delta_{sys}$ & \hspace{1.5cm}
 $\tilde{a}_\tau$   & \hspace{1.7cm} $|\tilde{d}_\tau(e cm)|$ \\
\hline
10  &\hspace{1.2cm}  $0\%$   &\hspace{1.2cm} [-0.00351; 0.00308]    & \hspace{1.5cm}   $1.813\times 10^{-17}$   \\
10  &\hspace{1.2cm}  $3\%$   &\hspace{1.2cm} [-0.00375; 0.00332]    & \hspace{1.5cm}   $1.977\times 10^{-17}$   \\
10  &\hspace{1.2cm}  $5\%$   &\hspace{1.2cm} [-0.00449; 0.00405]    & \hspace{1.5cm}   $2.385\times 10^{-17}$   \\
\hline
50   &\hspace{1.2cm} $0\%$   &\hspace{1.2cm} [-0.00245; 0.00203]    & \hspace{1.5cm}   $1.213\times 10^{-17}$  \\
50  &\hspace{1.2cm}  $3\%$   &\hspace{1.2cm} [-0.00348; 0.00305]    & \hspace{1.5cm}   $1.829\times 10^{-17}$  \\
50  &\hspace{1.2cm}  $5\%$  &\hspace{1.2cm}  [-0.00435; 0.00391]    & \hspace{1.5cm}   $2.306\times 10^{-17}$  \\
\hline
100   &\hspace{1.2cm} $0\%$  &\hspace{1.2cm} [-0.00211; 0.00169]    & \hspace{1.5cm}   $1.020\times 10^{-17}$  \\
100  &\hspace{1.2cm}  $3\%$  &\hspace{1.2cm} [-0.00344; 0.00301]    & \hspace{1.5cm}   $1.807\times 10^{-17}$  \\
100  &\hspace{1.2cm}  $5\%$ &\hspace{1.2cm}  [-0.00433; 0.00389]    & \hspace{1.5cm}   $2.296\times 10^{-17}$  \\
\hline
300   &\hspace{1.2cm} $0\%$   &\hspace{1.2cm} [-0.00169; 0.00127]    & \hspace{1.5cm}  $7.758\times 10^{-18}$  \\
300  &\hspace{1.2cm}  $3\%$   &\hspace{1.2cm} [-0.00342; 0.00299]    & \hspace{1.5cm}  $1.792\times 10^{-17}$  \\
300  &\hspace{1.2cm}  $5\%$  &\hspace{1.2cm}  [-0.00432; 0.00388]    & \hspace{1.5cm}  $2.289\times 10^{-17}$  \\
\hline
500   &\hspace{1.2cm} $0\%$  &\hspace{1.2cm} [-0.00153; 0.00111]    & \hspace{1.5cm}   $6.828\times 10^{-18}$  \\
500  &\hspace{1.2cm}  $3\%$  &\hspace{1.2cm} [-0.00341; 0.00298]    & \hspace{1.5cm}   $1.789\times 10^{-17}$  \\
500  &\hspace{1.2cm}  $5\%$ &\hspace{1.2cm}  [-0.00432; 0.00388]    & \hspace{1.5cm}   $2.288\times 10^{-17}$  \\
\hline\hline
\end{tabular}
\end{center}
\end{table}

\begin{table}[!ht]
\caption{Sensitivity on the $\tilde{a}_\tau$ magnetic moment and the $\tilde{d}_\tau$ electric dipole moment for $\sqrt{s}=1.5\hspace{0.8mm}TeV$
and ${\cal L}=100, 300, 500, 1000, 1500\hspace{0.8mm}fb^{-1}$ at $95\%$ C.L. through the process $\gamma\gamma \to \tau \bar\tau\gamma$.}
\begin{center}
 \begin{tabular}{ccccc}
\hline\hline
\multicolumn{4}{c}{$\sqrt{s}=1.5\hspace{0.8mm}TeV$, \hspace{5mm}  $95\%$ C.L.}\\
 \hline
 \cline{1-4}  ${\cal L}\hspace{0.8mm}(fb^{-1})$  & \hspace{1.5cm} $\delta_{sys}$ & \hspace{1.5cm}
 $\tilde{a}_\tau$   & \hspace{1.7cm} $|\tilde{d}_\tau(e cm)|$ \\
\hline
100  &\hspace{1.2cm}  $0\%$   &\hspace{1.2cm} [-0.00096; 0.00088]    & \hspace{1.5cm}       $5.121\times 10^{-18}$   \\
100  &\hspace{1.2cm}  $3\%$   &\hspace{1.2cm} [-0.00142; 0.00134]    & \hspace{1.5cm}       $7.695\times 10^{-18}$   \\
100  &\hspace{1.2cm}  $5\%$   &\hspace{1.2cm} [-0.00179; 0.00171]    & \hspace{1.5cm}       $9.756\times 10^{-18}$   \\
\hline
300   &\hspace{1.2cm} $0\%$   &\hspace{1.2cm} [-0.00074; 0.00066]    & \hspace{1.5cm}       $3.926\times 10^{-18}$  \\
300  &\hspace{1.2cm}  $3\%$   &\hspace{1.2cm} [-0.00140; 0.00132]    & \hspace{1.5cm}       $7.582\times 10^{-18}$  \\
300  &\hspace{1.2cm}  $5\%$   &\hspace{1.2cm} [-0.00178; 0.00170]    & \hspace{1.5cm}       $9.702\times 10^{-18}$  \\
\hline
500   &\hspace{1.2cm} $0\%$  &\hspace{1.2cm} [-0.00066; 0.00058]    & \hspace{1.5cm}        $3.475\times 10^{-18}$  \\
500  &\hspace{1.2cm}  $3\%$  &\hspace{1.2cm} [-0.00140; 0.00131]    & \hspace{1.5cm}        $7.559\times 10^{-18}$  \\
500  &\hspace{1.2cm}  $5\%$  &\hspace{1.2cm} [-0.00178; 0.00170]    & \hspace{1.5cm}        $9.691\times 10^{-18}$  \\
\hline
1000   &\hspace{1.2cm} $0\%$   &\hspace{1.2cm} [-0.00057; 0.00049]    & \hspace{1.5cm}      $2.952\times 10^{-18}$  \\
1000  &\hspace{1.2cm}  $3\%$   &\hspace{1.2cm} [-0.00139; 0.00131]    & \hspace{1.5cm}      $7.541\times 10^{-18}$  \\
1000  &\hspace{1.2cm}  $5\%$   &\hspace{1.2cm} [-0.00178; 0.00170]    & \hspace{1.5cm}      $9.683\times 10^{-18}$  \\
\hline
1500   &\hspace{1.2cm} $0\%$  &\hspace{1.2cm} [-0.00052; 0.00044]    & \hspace{1.5cm}       $2.688\times 10^{-18}$  \\
1500  &\hspace{1.2cm}  $3\%$  &\hspace{1.2cm} [-0.00139; 0.00131]    & \hspace{1.5cm}       $7.535\times 10^{-18}$  \\
1500  &\hspace{1.2cm}  $5\%$  &\hspace{1.2cm} [-0.00178; 0.00170]    & \hspace{1.5cm}       $9.680\times 10^{-18}$  \\
\hline\hline
\end{tabular}
\end{center}
\end{table}

\begin{table}[!ht]
\caption{Sensitivity on the $\tilde{a}_\tau$ magnetic moment and the $\tilde{d}_\tau$ electric dipole moment for $\sqrt{s}=3\hspace{0.8mm}TeV$
and ${\cal L}=100, 500, 1000, 20000, 3000\hspace{0.8mm}fb^{-1}$ at $95\%$ C.L. through the process $\gamma\gamma \to \tau \bar\tau\gamma$.}
\begin{center}
 \begin{tabular}{ccccc}
\hline\hline
\multicolumn{4}{c}{$\sqrt{s}=3\hspace{0.8mm}TeV$, \hspace{5mm}  $95\%$ C.L.}\\
 \hline
 \cline{1-4} ${\cal L}\hspace{0.8mm}(fb^{-1})$  & \hspace{1.5cm} $\delta_{sys}$ & \hspace{1.5cm} $\tilde{a}_\tau$   & \hspace{1.7cm}
 $|\tilde{d}_\tau(e cm)|$ \\
\hline
100  &\hspace{1.2cm}  $0\%$   &\hspace{1.2cm} [-0.00070; 0.00060]    & \hspace{1.5cm}    $3.612\times 10^{-18}$   \\
100  &\hspace{1.2cm}  $3\%$   &\hspace{1.2cm} [-0.00087; 0.00078]    & \hspace{1.5cm}    $4.591\times 10^{-18}$   \\
100  &\hspace{1.2cm}  $5\%$   &\hspace{1.2cm} [-0.00108; 0.00098]    & \hspace{1.5cm}    $5.740\times 10^{-18}$   \\
\hline
500   &\hspace{1.2cm} $0\%$   &\hspace{1.2cm} [-0.00048; 0.00039]    & \hspace{1.5cm}    $2.418\times 10^{-18}$  \\
500  &\hspace{1.2cm}  $3\%$   &\hspace{1.2cm} [-0.00084; 0.00075]    & \hspace{1.5cm}    $4.422\times 10^{-18}$  \\
500  &\hspace{1.2cm}  $5\%$  &\hspace{1.2cm}  [-0.00106; 0.00097]    & \hspace{1.5cm}    $5.658\times 10^{-18}$  \\
\hline
1000   &\hspace{1.2cm} $0\%$  &\hspace{1.2cm} [-0.00041; 0.00032]    & \hspace{1.5cm}    $2.034\times 10^{-18}$  \\
1000  &\hspace{1.2cm}  $3\%$  &\hspace{1.2cm} [-0.00084; 0.00074]    & \hspace{1.5cm}    $4.399\times 10^{-18}$  \\
1000  &\hspace{1.2cm}  $5\%$ &\hspace{1.2cm}  [-0.00106; 0.00097]    & \hspace{1.5cm}    $5.647\times 10^{-18}$  \\
\hline
2000   &\hspace{1.2cm} $0\%$   &\hspace{1.2cm} [-0.00035; 0.00026]    & \hspace{1.5cm}   $1.711\times 10^{-18}$  \\
2000  &\hspace{1.2cm}  $3\%$   &\hspace{1.2cm} [-0.00083; 0.00074]    & \hspace{1.5cm}   $4.388\times 10^{-18}$  \\
2000  &\hspace{1.2cm}  $5\%$  &\hspace{1.2cm}  [-0.00106; 0.00097]    & \hspace{1.5cm}   $5.642\times 10^{-18}$  \\
\hline
3000   &\hspace{1.2cm} $0\%$  &\hspace{1.2cm} [-0.00033; 0.00023]    & \hspace{1.5cm}    $1.546\times 10^{-18}$  \\
3000  &\hspace{1.2cm}  $3\%$  &\hspace{1.2cm} [-0.00083; 0.00074]    & \hspace{1.5cm}    $4.384\times 10^{-18}$  \\
3000  &\hspace{1.2cm}  $5\%$ &\hspace{1.2cm}  [-0.00106; 0.00097]    & \hspace{1.5cm}    $5.640\times 10^{-18}$  \\
\hline\hline
\end{tabular}
\end{center}
\end{table}

\section{Conclusions}

In this paper we have presented a complete study of the total cross section as well as of the anomalous magnetic and electric
dipole moments of the tau-lepton in the scenarios $\gamma\gamma \to \tau^+\tau^-$ and $\gamma\gamma \to \tau^+\tau^-\gamma$ at the
CLIC. To investigate these scenarios, we have considered a set of parameters $\tilde{a}_\tau$ and $\tilde{d}_\tau$ for both processes, as well as the
parameters $\sqrt{s}$ and ${\cal L}$ of the collider. Furthermore, for each of these scenarios we considered a set of cuts appropriate
given by  Eqs. (12) and (24) to reduce the background and to optimize the signal sensitivity to the particles of the $\tau^+\tau^-
(\tau^+\tau^-\gamma)$ final state.

Overall, our study shows that the scenario $\gamma\gamma \to \tau^+\tau^-$ with two-tau in the final state, is more representative and
projects a better sensitivity in both, the total cross section and in the bounds on the electromagnetic dipole moments in comparison
with the process $\gamma\gamma \to \tau^+\tau^-\gamma$ in the entire range of center-of-mass energies and luminosities of the future CLIC.
Another important aspect in our study that is worth mentioning to distinguish the sensitivity in our results is the incorporation of the
systematic uncertainties of $\delta_{sys}=0, 3, 5\hspace{0.8mm}\%$.

In summary, we have shown that the two-photon $\gamma\gamma \to \tau^+\tau^-$ and $\gamma\gamma \to \tau^+\tau^-\gamma$ processes
at the CLIC leads to a remarkable improvement in the existing experimental bounds on the $\tilde{a}_\tau$ and $\tilde{d}_\tau$. In the case of the
scenario $\gamma\gamma \to \tau^+\tau^-$ we get a significant improvement of the order of magnitude of $3.466\times 10^2$ for
the upper bound and of $0.764\times 10^2$ for the lower bound in comparison with the results published by the DELPHI and BELLE Collaborations
for the reaction $e^+e^- \to \tau^+\tau^-$ \cite{Abdallah,Inami}. In the case of the scenario $\gamma\gamma \to \tau^+\tau^-\gamma$
the improvement is of the order of $2.060\times 10^2$ and $2.826\times 10^2$ for the upper and lower bounds, respectively, in comparison
with the reported by the L3 and OPAL Collaborations for the process $e^+e^- \to \tau^+\tau^-\gamma$ \cite{Acciarri,Ackerstaff}, as shown in
Table I. Our results indicate and shown that the processes $\gamma \gamma \to \tau^+ \tau^-$ and $\gamma\gamma \to \tau^+\tau^-\gamma$
are more suitable for probing the electromagnetic dipole moments of the $\tau$-lepton in the future $e^+e^-$ linear colliders such as the CLIC at the $\gamma\gamma$ mode.

\vspace*{2cm}

%\newpage

\begin{center}
{\bf Acknowledgments}
\end{center}%

A. G. R. and M. A. H. R acknowledges support from SNI and PROFOCIE (M\'exico).

\vspace*{3cm}

%\newpage

\pagebreak

\begin{figure}[H]
\centerline{\scalebox{0.55}{\includegraphics{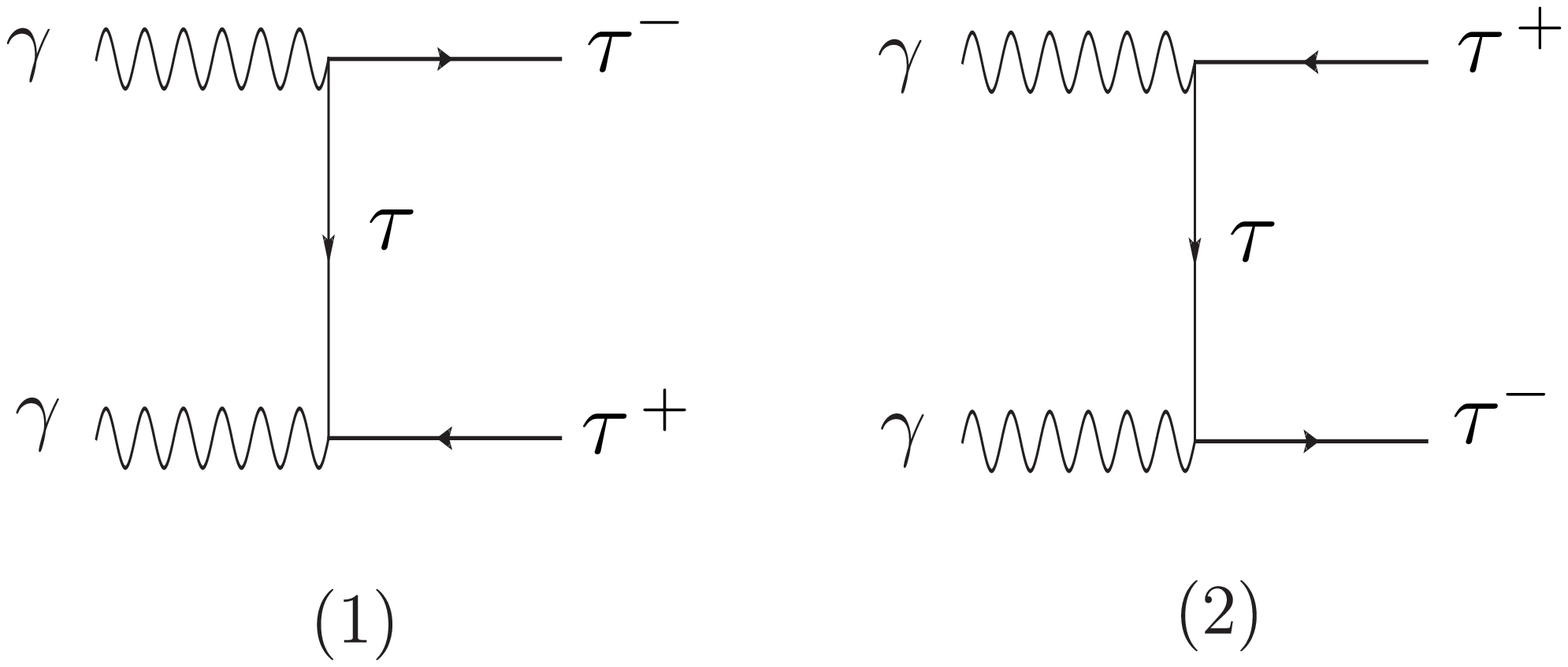}}}
\caption{ \label{fig:gamma1} The Feynman diagrams for the process
$\gamma\gamma \to \tau^+ \tau^-$.}
\label{Fig.1}
\end{figure}

\begin{figure}[H]
\centerline{\scalebox{0.55}{\includegraphics{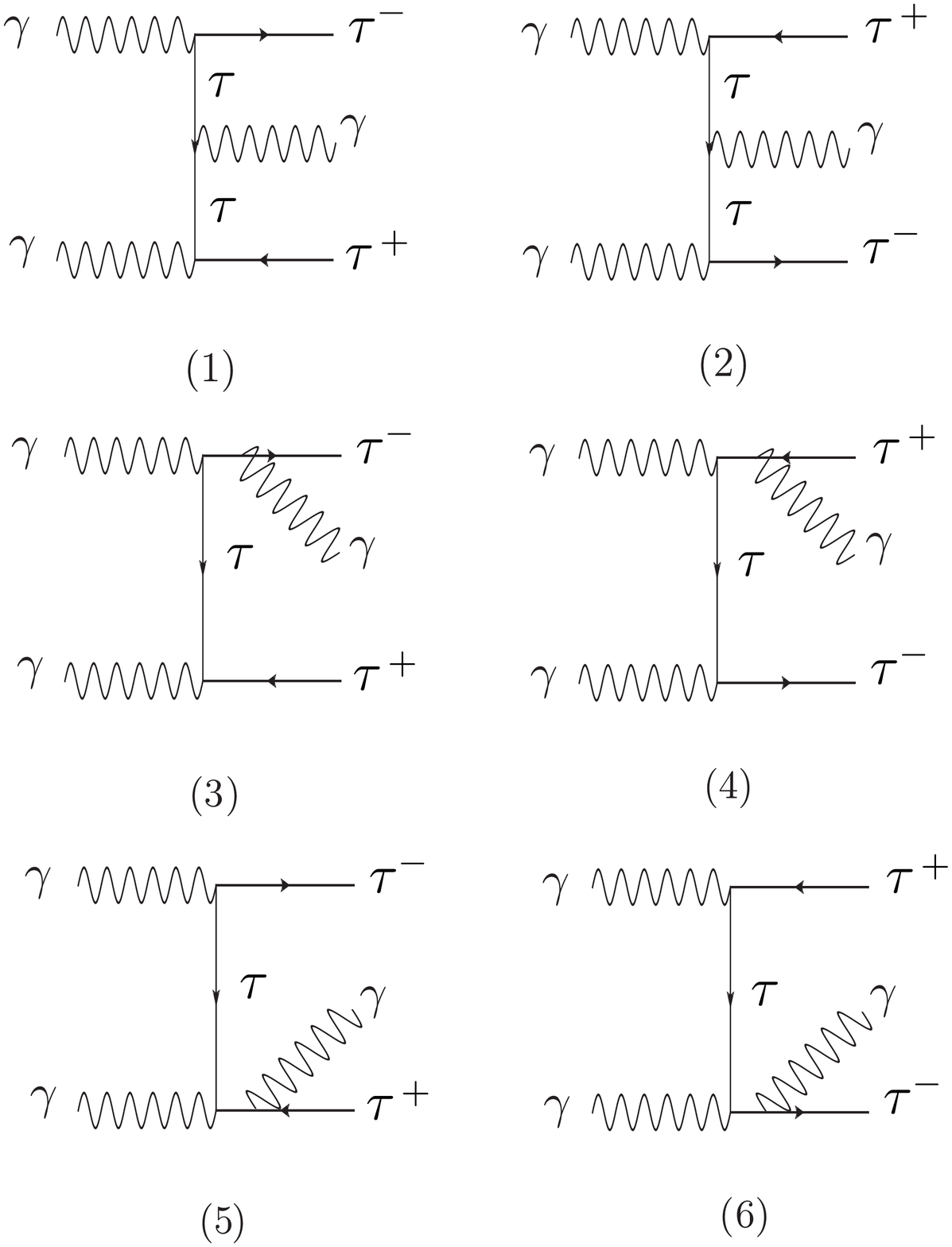}}}
\caption{ \label{fig:gamma2} The Feynman diagrams for the process
$\gamma\gamma \to \tau^+ \tau^-\gamma$.}
\label{Fig.2}
\end{figure}

\begin{figure}[H]
\centerline{\scalebox{1.5}{\includegraphics{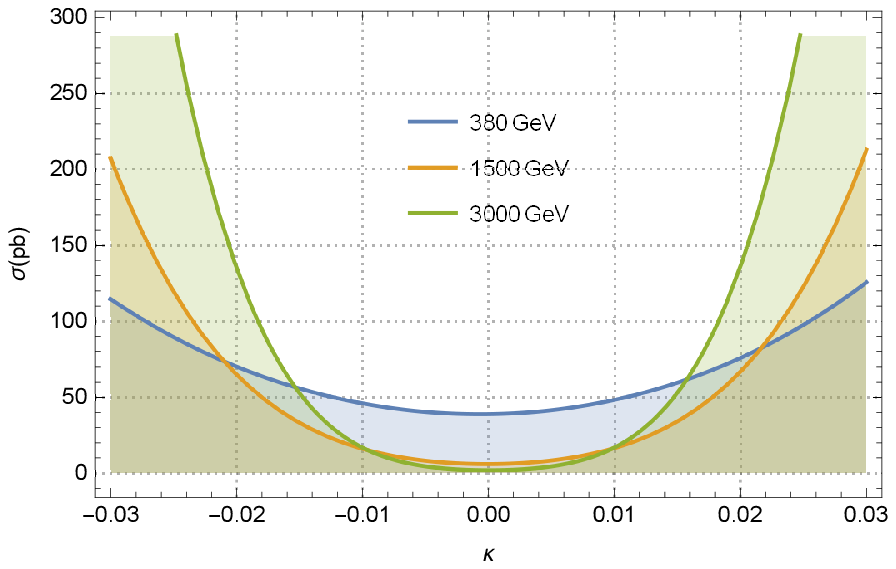}}}
\caption{ \label{fig:gamma1} The total cross sections of the process
$\gamma\gamma \rightarrow \tau^+ \tau^-$ as a function of $\kappa$
for center-of-mass energies of $\sqrt{s}=380, 1500, 3000$\hspace{0.8mm}$GeV$.}
\label{Fig.3}
\end{figure}

\begin{figure}[H]
\centerline{\scalebox{1.5}{\includegraphics{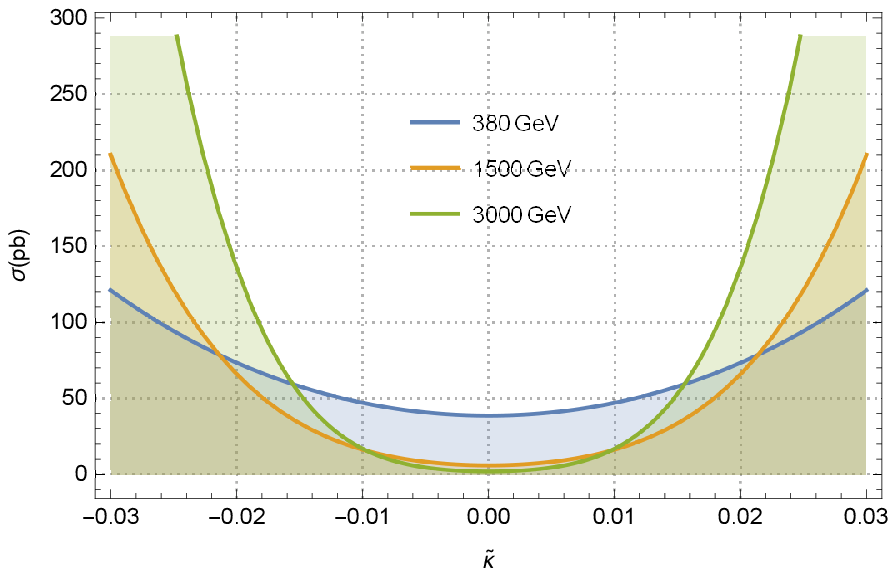}}}
\caption{ \label{fig:gamma2} Same as in Fig. 3, but for $\tilde{\kappa}$.}
\label{Fig.4}
\end{figure}

\begin{figure}[H]
\centerline{\scalebox{1}{\includegraphics{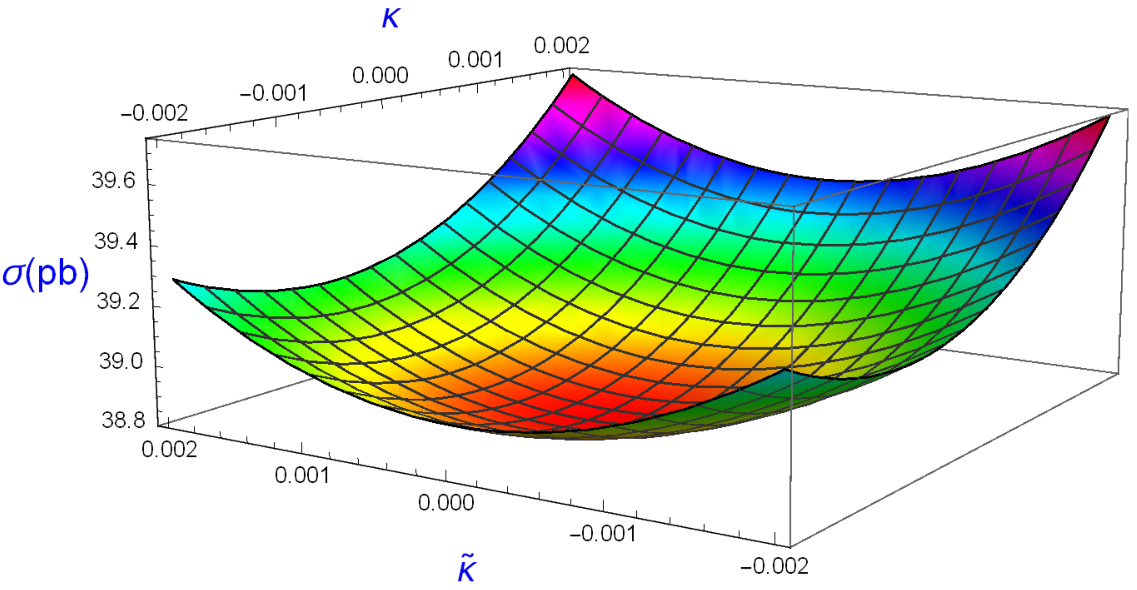}}}
\caption{ \label{fig:gamma3} The total cross sections of the process
$\gamma\gamma \rightarrow \tau^+ \tau^-$ as a function of $\kappa$ and $\tilde{\kappa}$
for center-of-mass energy of $\sqrt{s}=380\hspace{0.8mm}GeV$.}
\label{Fig.5}
\end{figure}

\begin{figure}[H]
\centerline{\scalebox{1}{\includegraphics{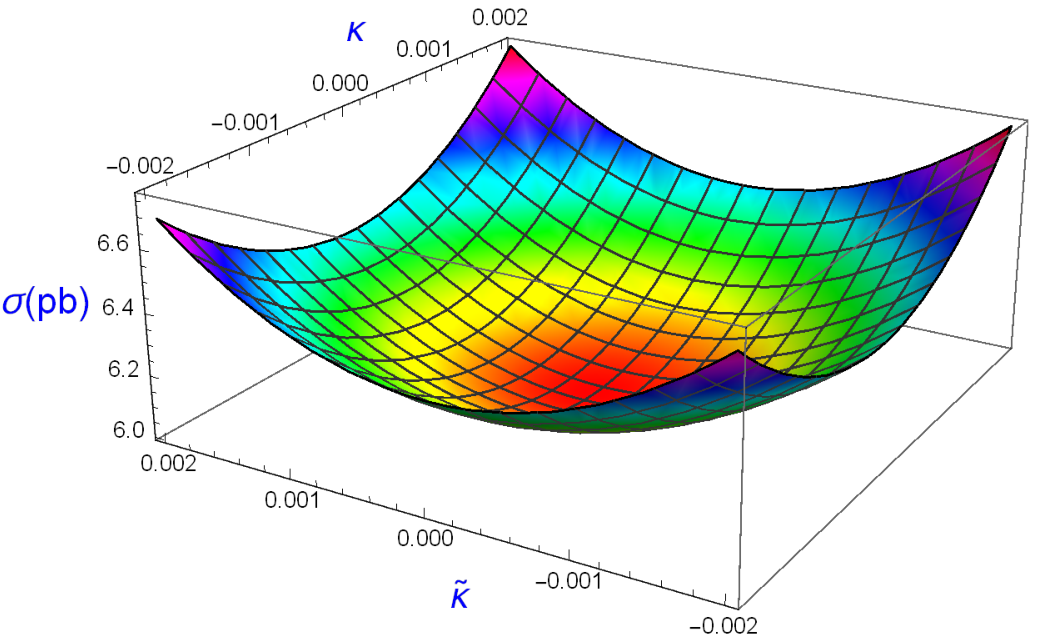}}}
\caption{ \label{fig:gamma15} Same as in Fig. 5, but for $\sqrt{s}=1500\hspace{0.8mm}GeV$.}
\label{Fig.6}
\end{figure}

\begin{figure}[H]
\centerline{\scalebox{1.2}{\includegraphics{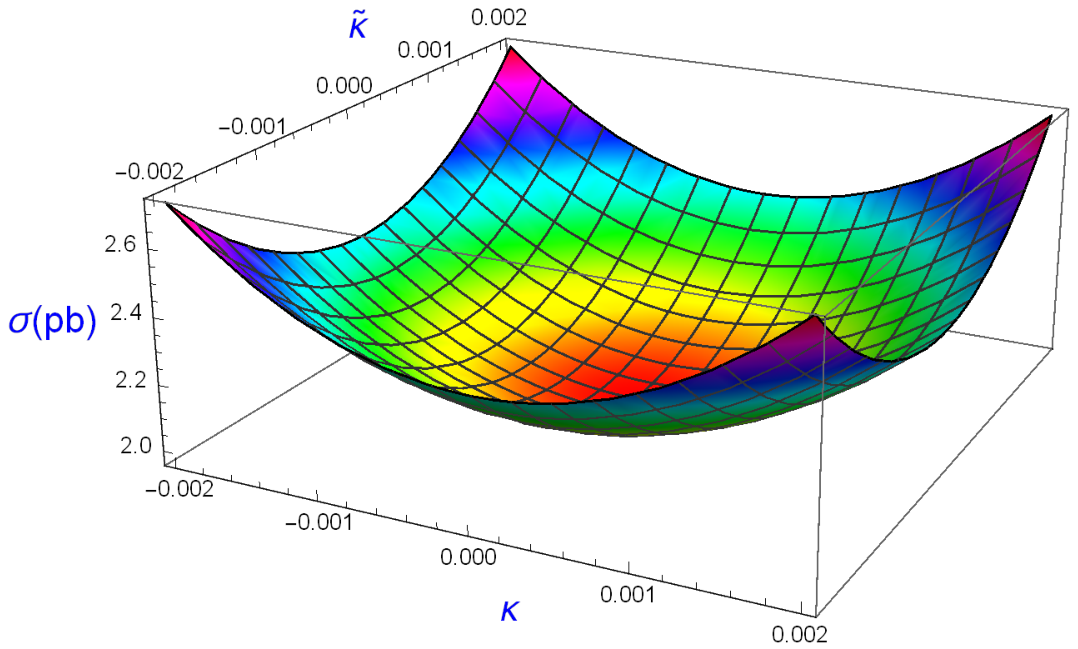}}}
\caption{ \label{fig:gamma15} Same as in Fig. 5, but for $\sqrt{s}=3000\hspace{0.8mm}GeV$.}
\label{Fig.6}
\end{figure}

\begin{figure}[H]
\centerline{\scalebox{1.2}{\includegraphics{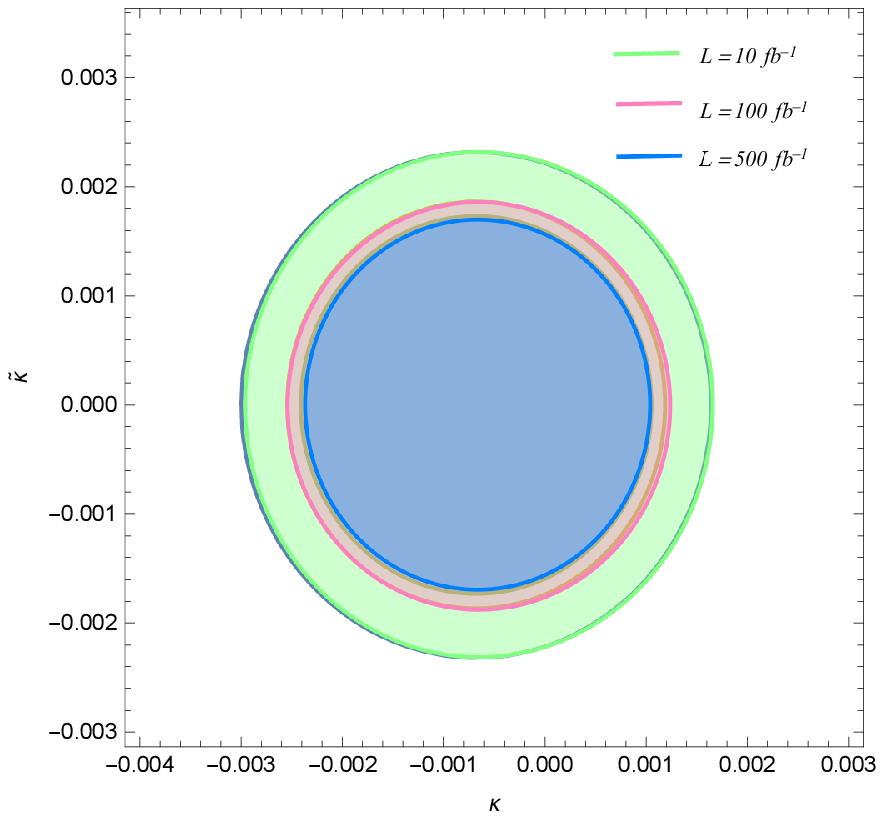}}}
\caption{ \label{fig:gamma15} Bounds contours at the $95\% \hspace{1mm}C.L.$ in the
$(\kappa -\tilde{\kappa})$ plane for the process $\gamma\gamma \rightarrow \tau^+ \tau^-$
with the ${\cal L}=10, 100, 500\hspace{0.8mm} fb^{-1}$ and for center-of-mass energy
of $\sqrt{s}=380\hspace{0.8mm}GeV$.}
\label{Fig.6}
\end{figure}

\begin{figure}[H]
\centerline{\scalebox{1.1}{\includegraphics{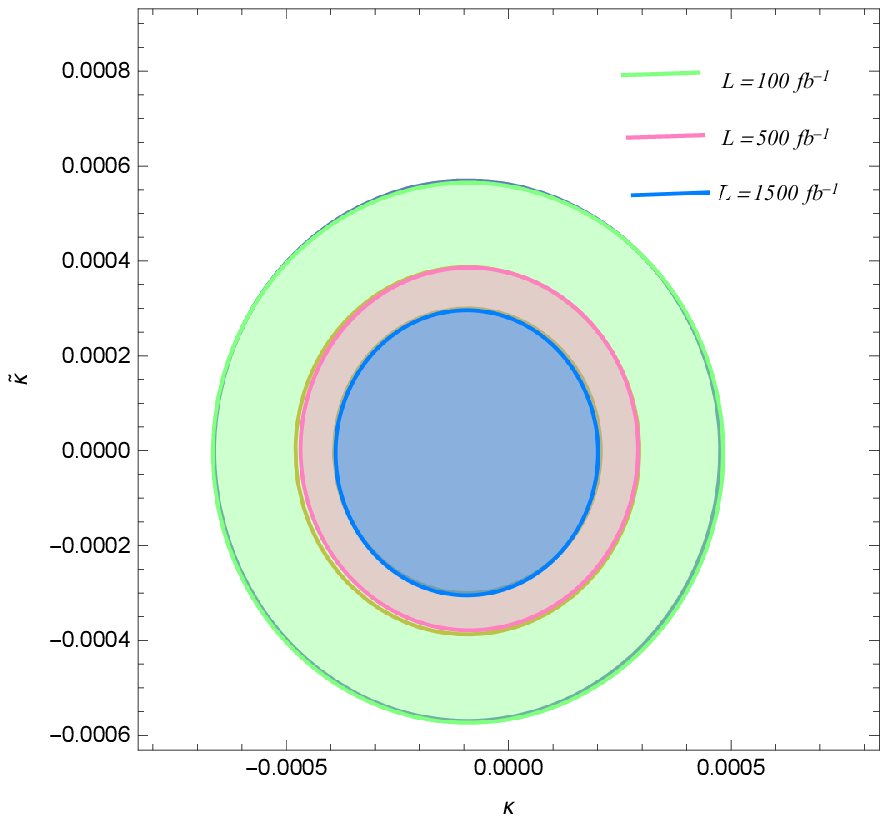}}}
\caption{ \label{fig:gamma15} Same as in Fig. 8, but for
${\cal L}=100, 500, 1500\hspace{0.8mm} fb^{-1}$ and for center-of-mass energy
of $\sqrt{s}=1500\hspace{0.8mm}GeV$.}
\label{Fig.6}
\end{figure}

\begin{figure}[H]
\centerline{\scalebox{1.1}{\includegraphics{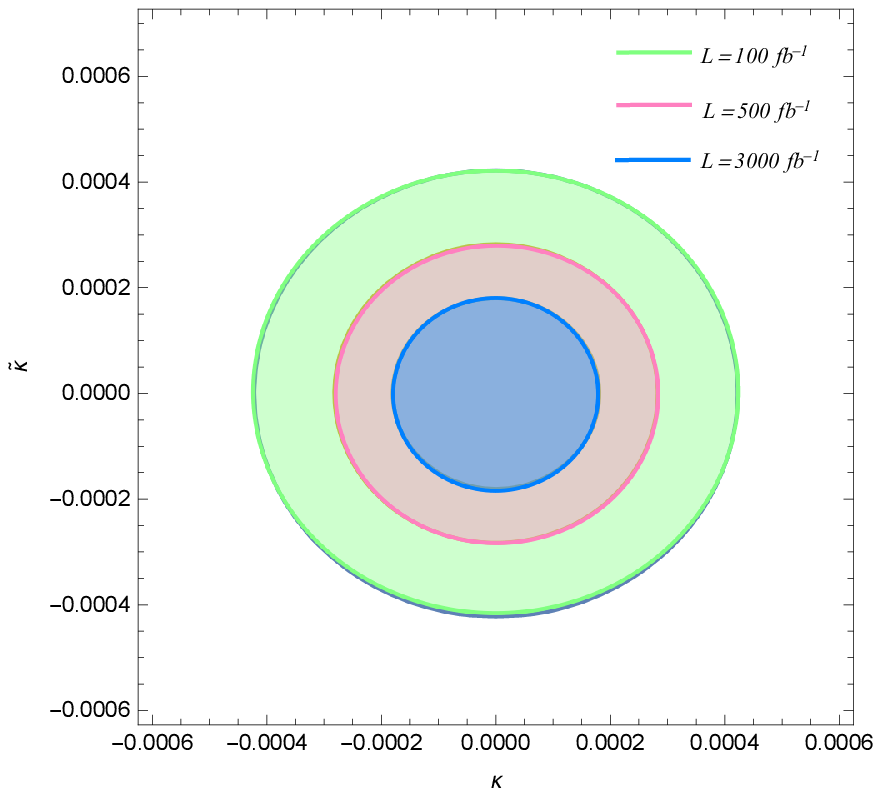}}}
\caption{ \label{fig:gamma15} Same as in Fig. 8, but for
${\cal L}=100, 500, 3000\hspace{0.8mm} fb^{-1}$ and for center-of-mass energy
of $\sqrt{s}=3000\hspace{0.8mm}GeV$.}
\label{Fig.6}
\end{figure}

\begin{figure}[H]
\centerline{\scalebox{0.9}{\includegraphics{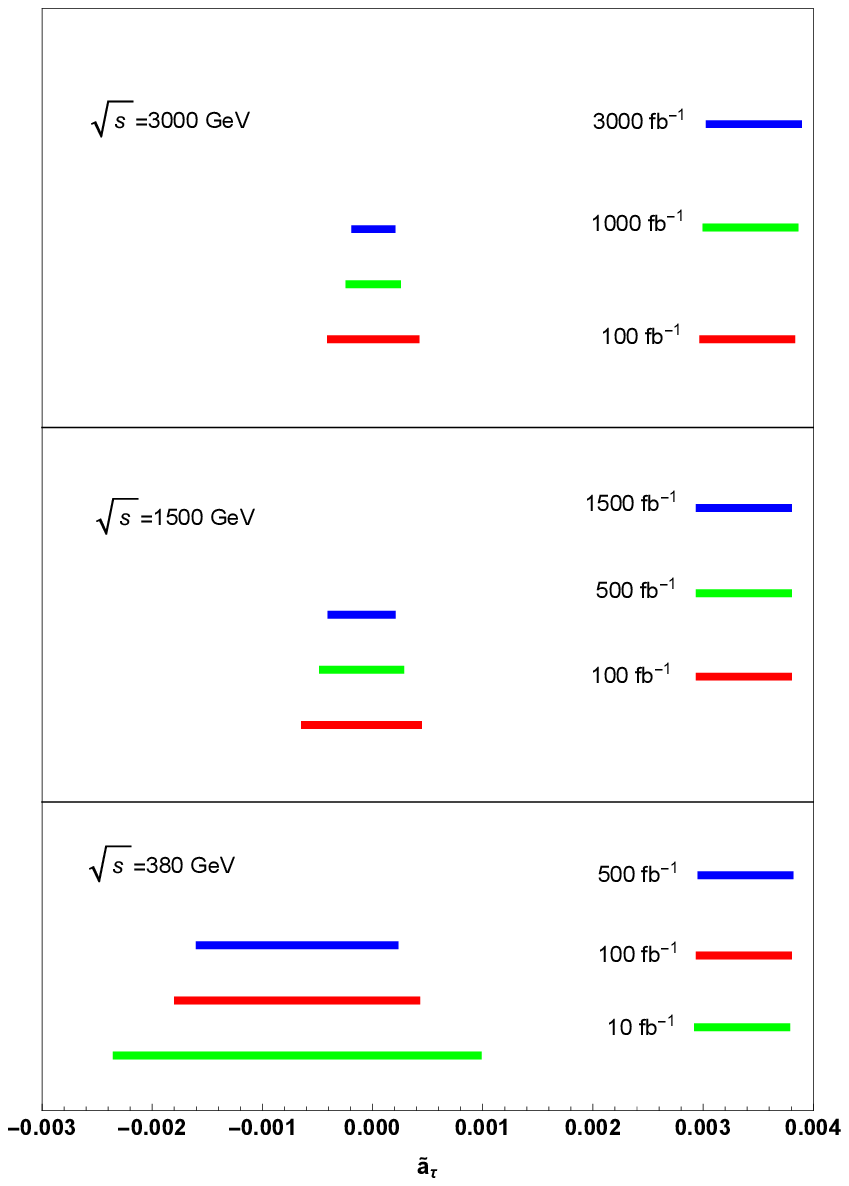}}}
\caption{ \label{fig:gamma15} Comparison of precisions on $\tilde{a}_\tau$ in the process
$\gamma\gamma \rightarrow \tau^+ \tau^-$ expected at the CLIC.
We assume luminosities of,  Top panel: ${\cal L}=100, 1000, 3000\hspace{0.8mm} fb^{-1}$ and $\sqrt{s}=3000\hspace{0.8mm}GeV$.
Central panel: ${\cal L}=100, 500, 1500\hspace{0.8mm} fb^{-1}$ and $\sqrt{s}=1500\hspace{0.8mm}GeV$.
Bottom panel: ${\cal L}=10, 100, 500\hspace{0.8mm} fb^{-1}$ and $\sqrt{s}=380\hspace{0.8mm}GeV$.}
\label{Fig.6}
\end{figure}

\begin{figure}[H]
\centerline{\scalebox{1}{\includegraphics{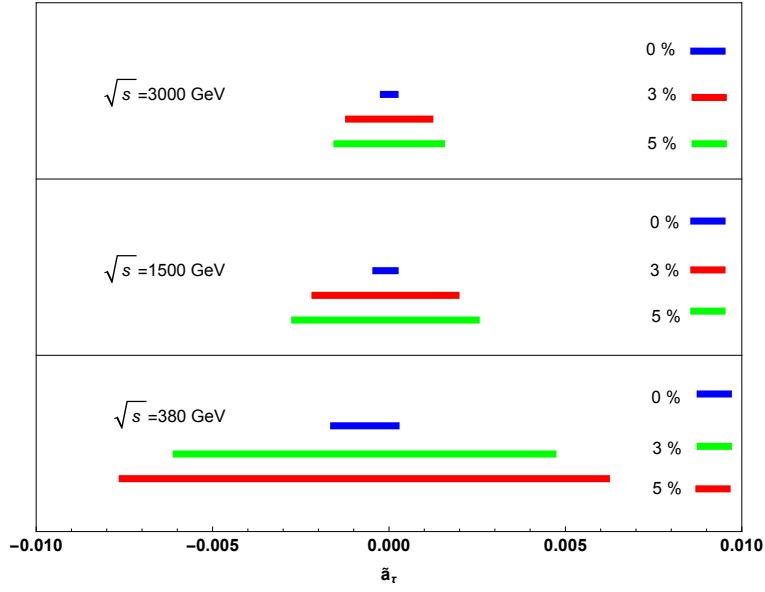}}}
\caption{ \label{fig:gamma15} Same as in Fig. 11, but for,
Top panel: $\delta_{sys}=0, 3, 5\hspace{0.8mm}\%$ with ${\cal L}=3000\hspace{0.8mm} fb^{-1}$ and $\sqrt{s}=3000\hspace{0.8mm}GeV$.
Central panel: $\delta_{sys}=0, 3, 5\hspace{0.8mm}\%$ with ${\cal L}= 1500\hspace{0.8mm} fb^{-1}$ and $\sqrt{s}=1500\hspace{0.8mm}GeV$.
Bottom panel: $\delta_{sys}=0, 3, 5\hspace{0.8mm}\%$ with ${\cal L}=500\hspace{0.8mm} fb^{-1}$ and $\sqrt{s}=380\hspace{0.8mm}GeV$.}
\label{Fig.6}
\end{figure}

\begin{figure}[H]
\centerline{\scalebox{1.3}{\includegraphics{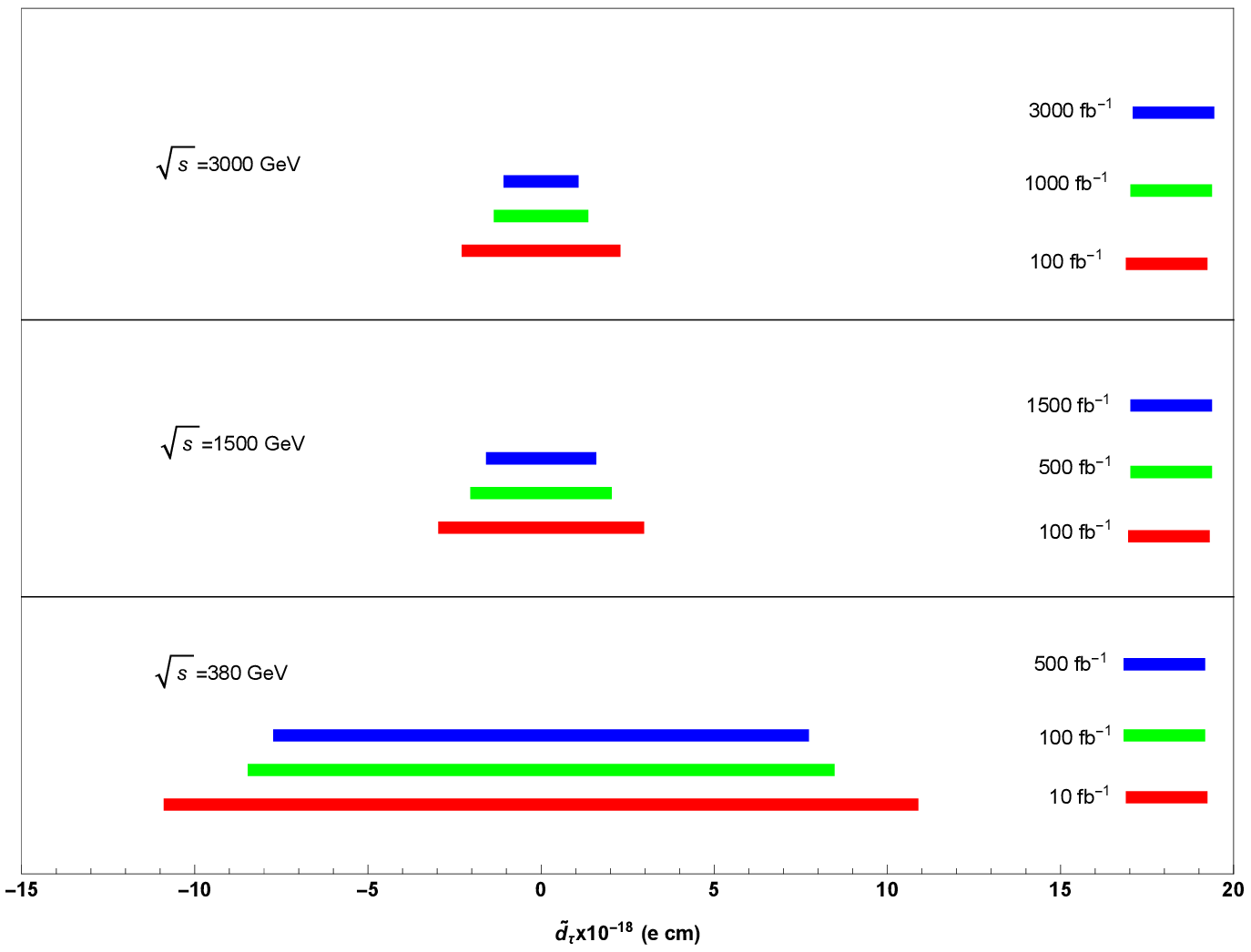}}}
\caption{ \label{fig:gamma15} Same as in Fig. 11, but for $\tilde{d}_\tau$.}
\label{Fig.6}
\end{figure}

\begin{figure}[H]
\centerline{\scalebox{0.8}{\includegraphics{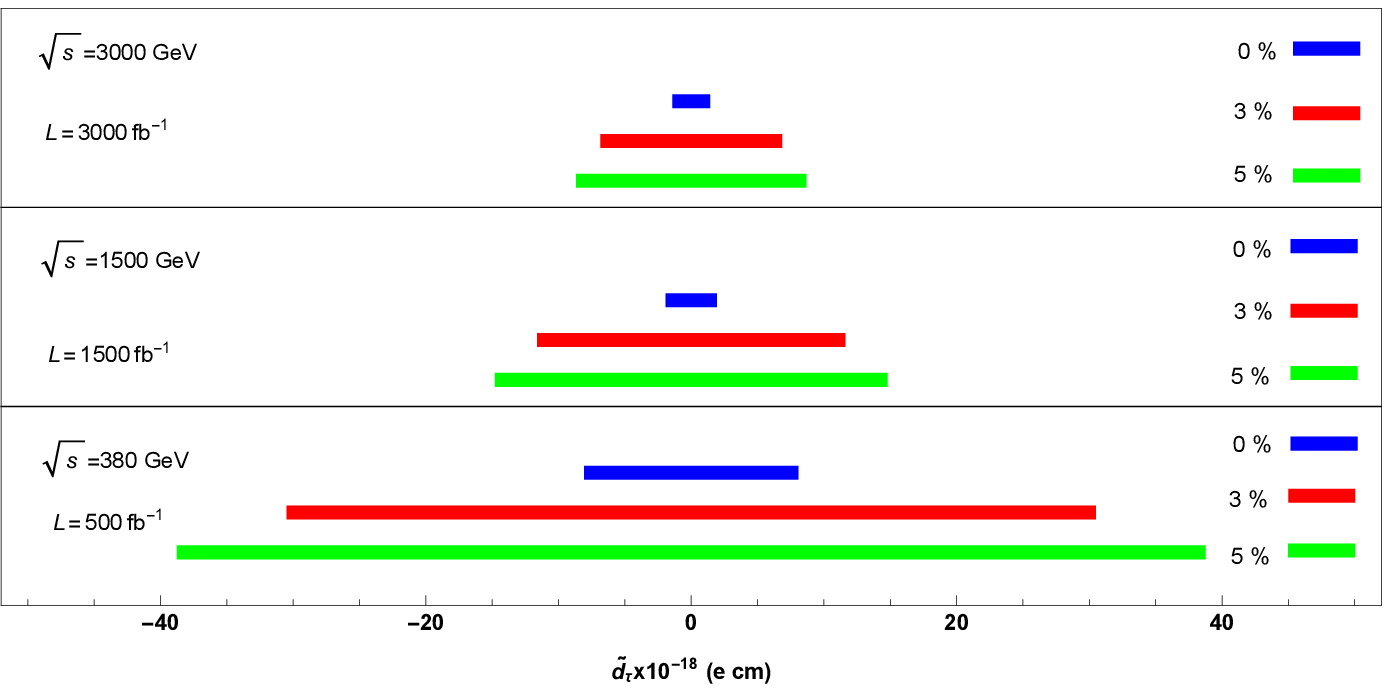}}}
\caption{ \label{fig:gamma15} Same as in Fig. 12, but for $\tilde{d}_\tau$.}
\label{Fig.6}
\end{figure}

\begin{figure}[H]
\centerline{\scalebox{1.5}{\includegraphics{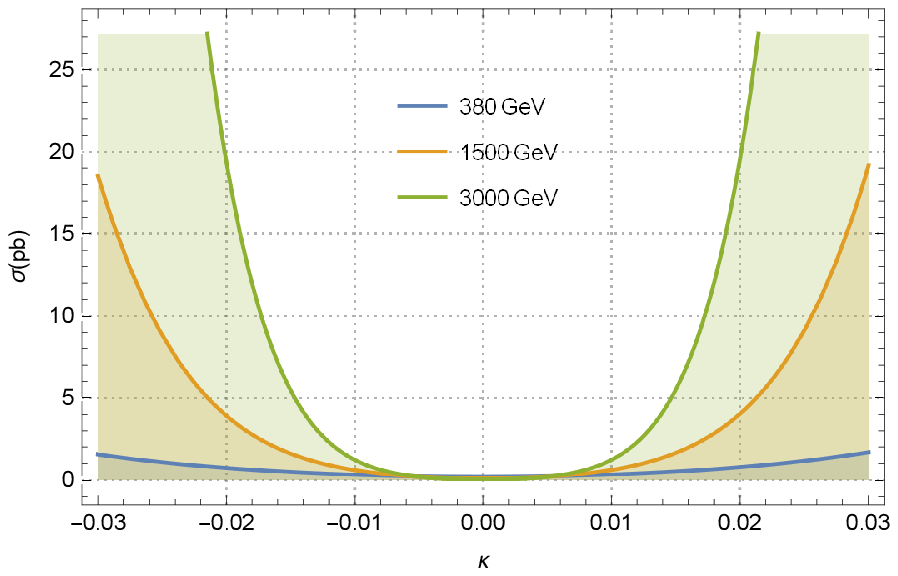}}}
\caption{ \label{fig:gamma15} The total cross sections of the process
$\gamma\gamma \rightarrow \tau^+ \tau^-\gamma$ as a function of $\kappa$
for center-of-mass energies of $\sqrt{s}=380, 1500, 3000$\hspace{0.8mm}$GeV$.}
\label{Fig.6}
\end{figure}

\begin{figure}[H]
\centerline{\scalebox{1.5}{\includegraphics{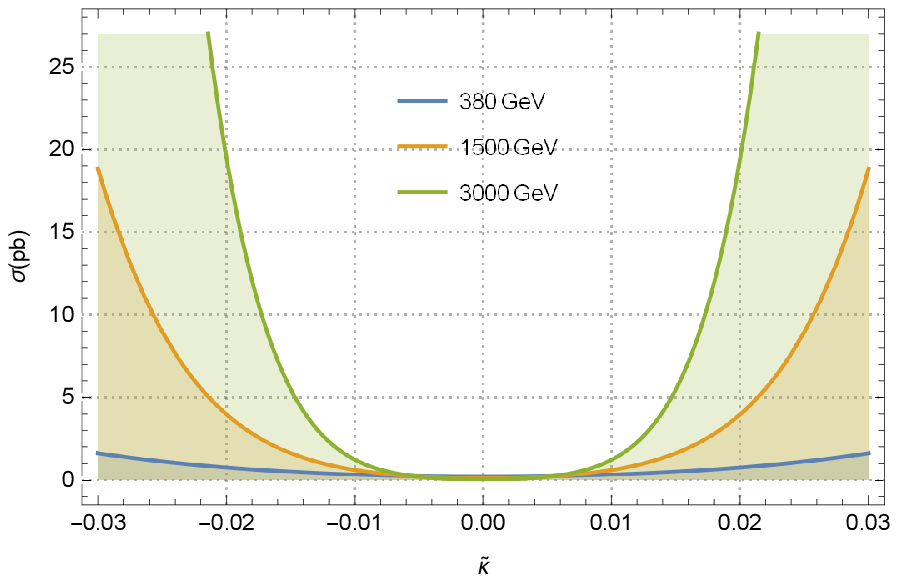}}}
\caption{ \label{fig:gamma15} Same as in Fig. 15, but for $\tilde{\kappa}$.}
\label{Fig.6}
\end{figure}

\begin{figure}[H]
\centerline{\scalebox{1.25}{\includegraphics{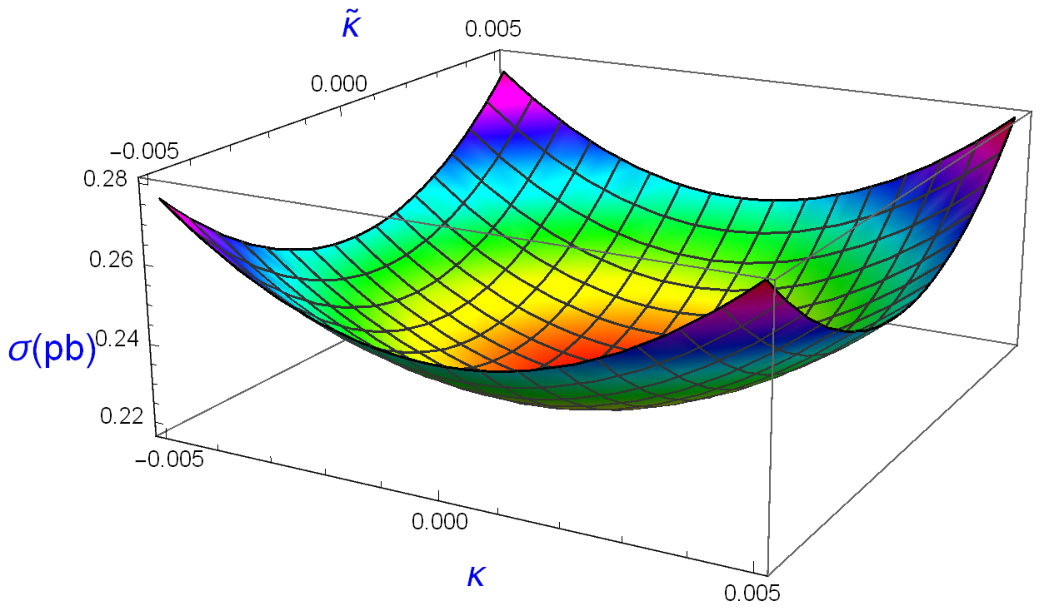}}}
\caption{ \label{fig:gamma15} The total cross sections of the process
$\gamma\gamma \rightarrow \tau^+ \tau^-\gamma$ as a function of $\kappa$ and $\tilde{\kappa}$
for center-of-mass energy of $\sqrt{s}=380\hspace{0.8mm}GeV$.}
\label{Fig.6}
\end{figure}

\begin{figure}[H]
\centerline{\scalebox{1.1}{\includegraphics{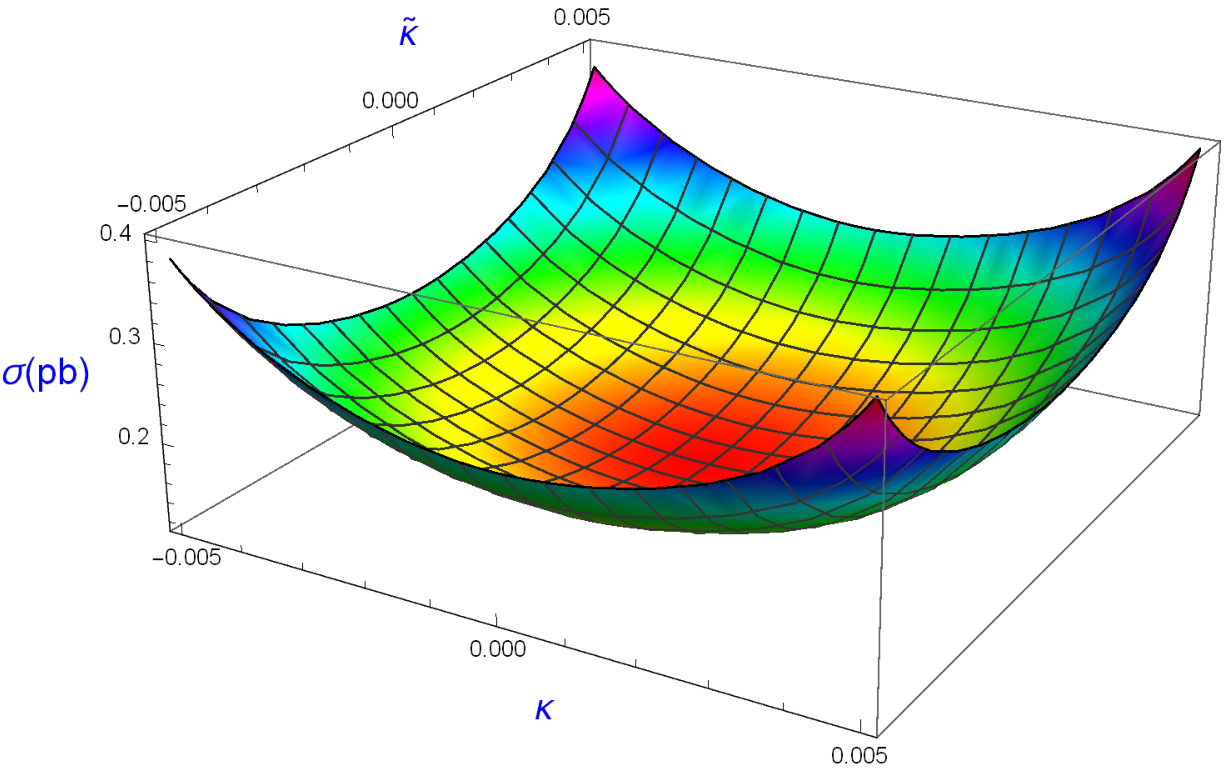}}}
\caption{ \label{fig:gamma15} Same as in Fig. 17, but for
 $\sqrt{s}=1500\hspace{0.8mm}GeV$.}
\label{Fig.6}
\end{figure}

\begin{figure}[H]
\centerline{\scalebox{0.9}{\includegraphics{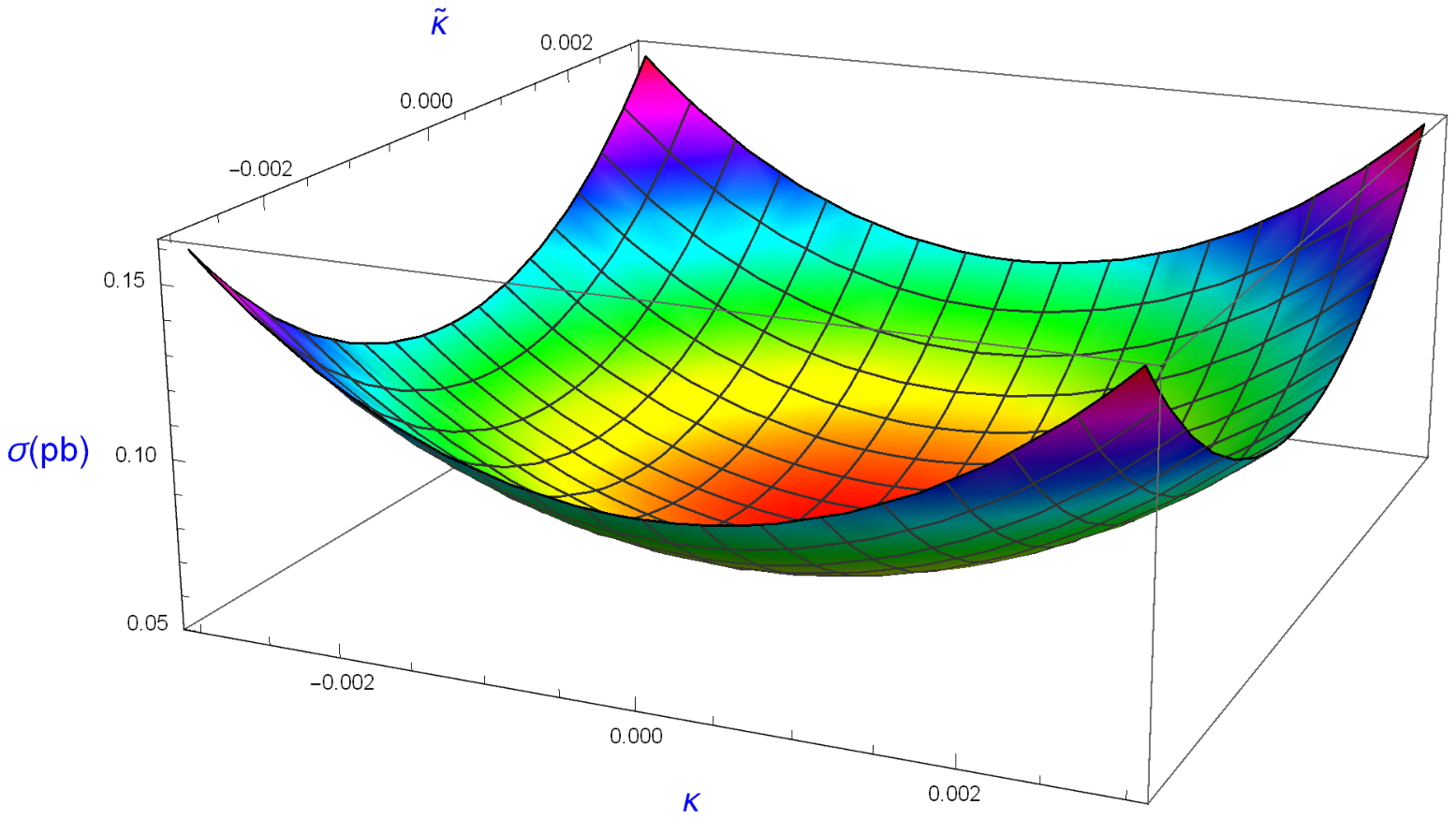}}}
\caption{ \label{fig:gamma15} Same as in Fig. 17, but for
$\sqrt{s}=3000\hspace{0.8mm}GeV$.}
\label{Fig.6}
\end{figure}

\begin{figure}[H]
\centerline{\scalebox{1.2}{\includegraphics{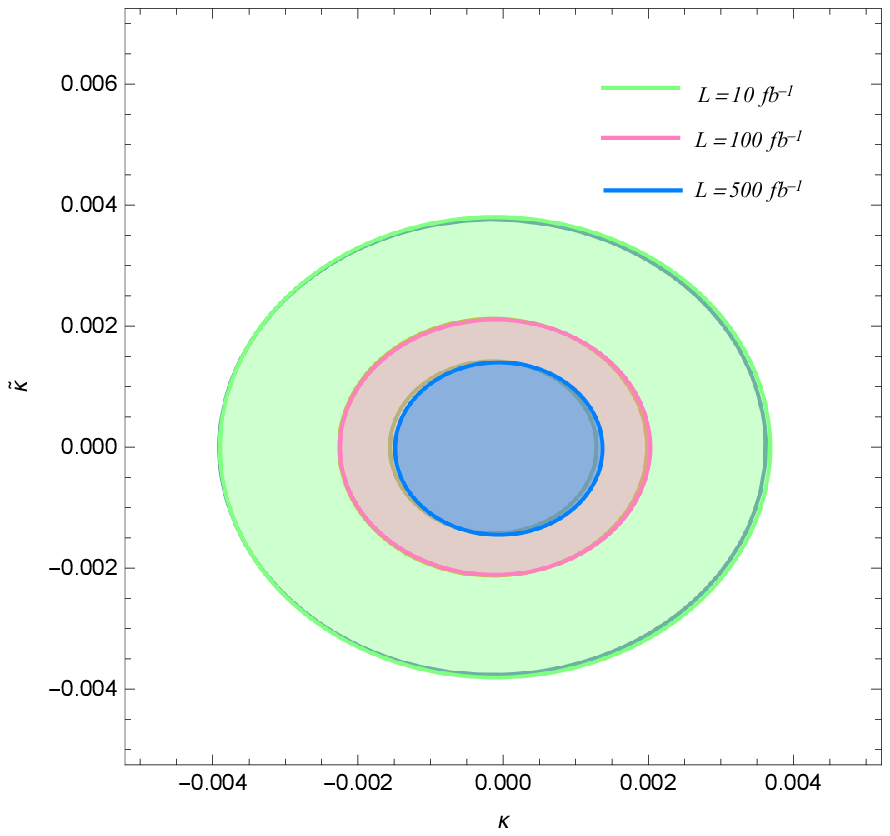}}}
\caption{ \label{fig:gamma15} Bounds contours at the $95\% \hspace{1mm}C.L.$ in the
$(\kappa -\tilde{\kappa})$ plane for the process $\gamma\gamma \rightarrow \tau^+ \tau^-\gamma$
with the ${\cal L}=10, 100, 500\hspace{0.8mm} fb^{-1}$ and for center-of-mass energy
of $\sqrt{s}=380\hspace{0.8mm}GeV$.}
\label{Fig.6}
\end{figure}

\begin{figure}[H]
\centerline{\scalebox{1.1}{\includegraphics{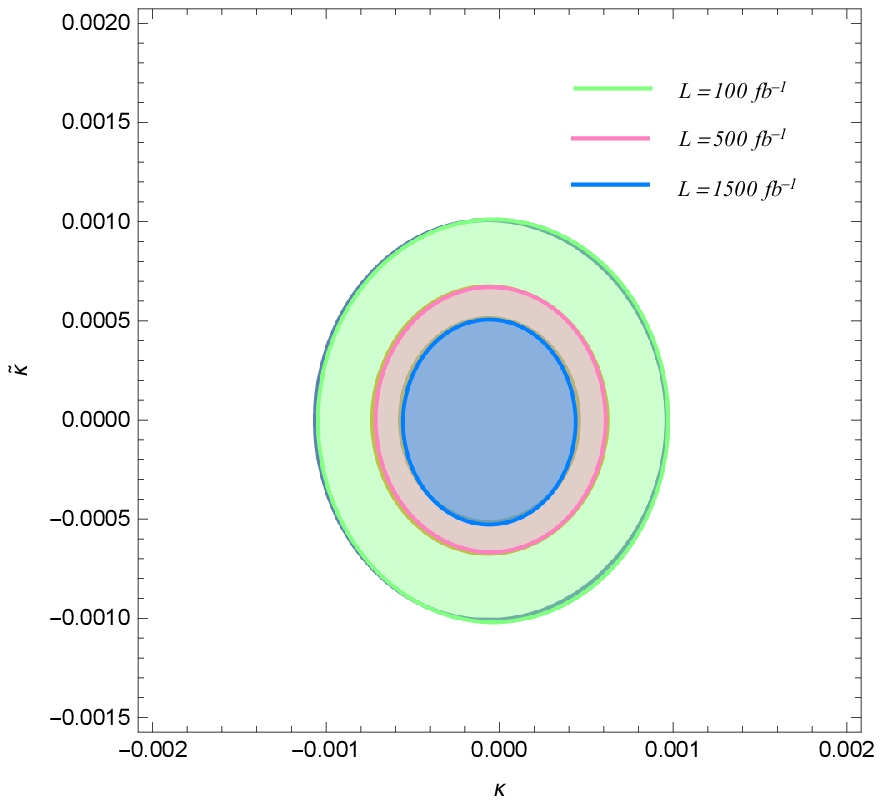}}}
\caption{ \label{fig:gamma15} Same as in Fig. 20, but for
${\cal L}=100, 500, 1500\hspace{0.8mm} fb^{-1}$ and for center-of-mass energy
of $\sqrt{s}=1500\hspace{0.8mm}GeV$.}
\label{Fig.6}
\end{figure}

\begin{figure}[H]
\centerline{\scalebox{1.1}{\includegraphics{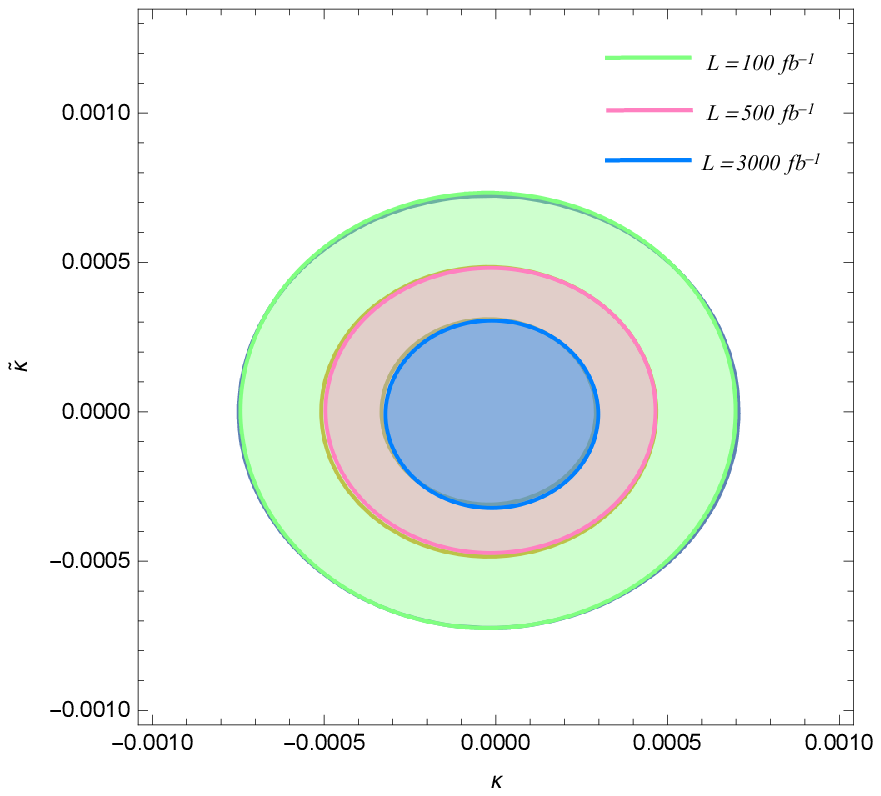}}}
\caption{ \label{fig:gamma15} Same as in Fig. 20, but for
${\cal L}=100, 500, 3000\hspace{0.8mm} fb^{-1}$ and for center-of-mass energy
of $\sqrt{s}=3000\hspace{0.8mm}GeV$.}
\label{Fig.6}
\end{figure}

\begin{figure}[H]
\centerline{\scalebox{0.8}{\includegraphics{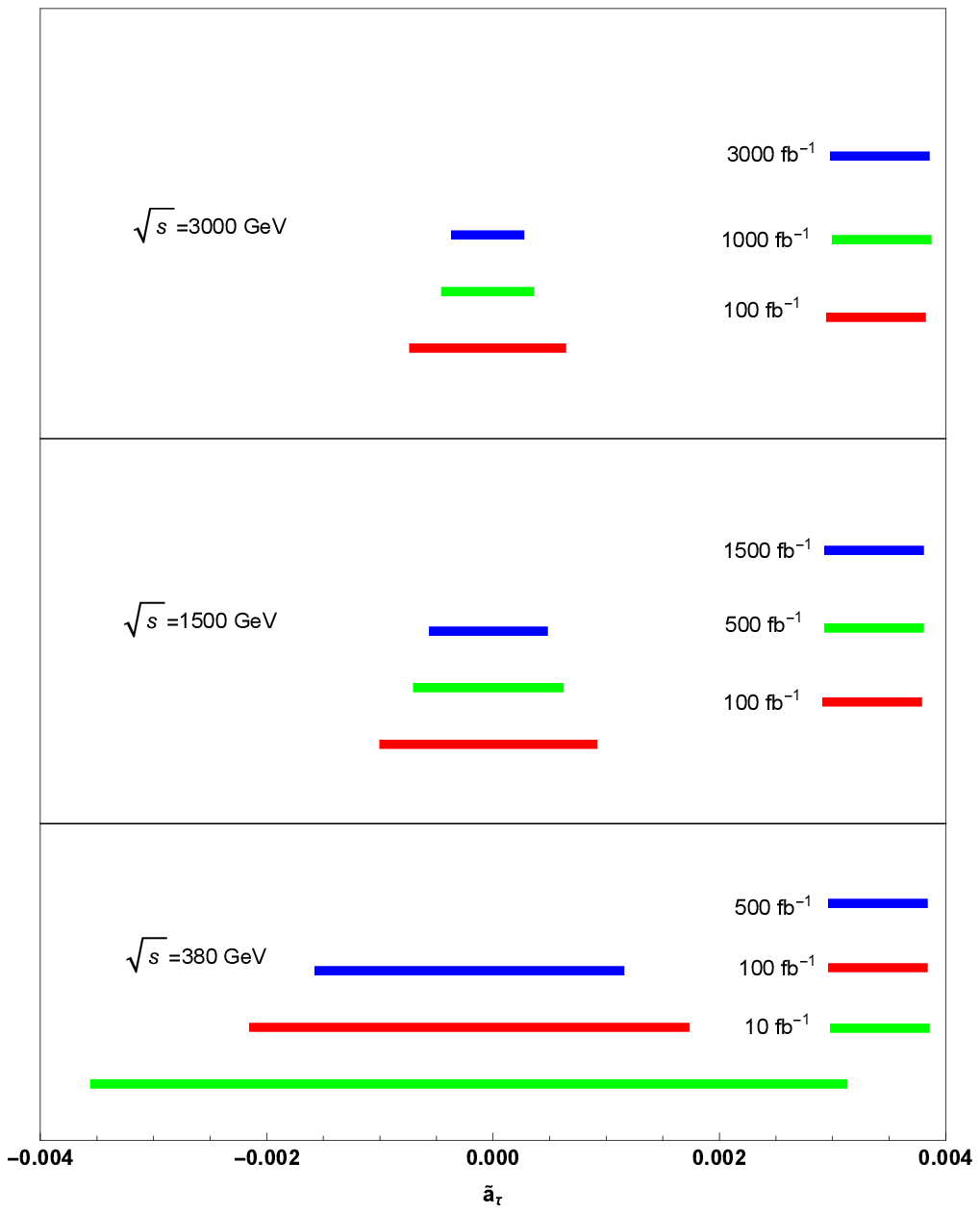}}}
\caption{ \label{fig:gamma15} Comparison of precisions on $\tilde{a}_\tau$ in the process
$\gamma\gamma \rightarrow \tau^+ \tau^-\gamma$ expected at the CLIC.
We assume luminosities of, Top panel: ${\cal L}=100, 1000, 3000\hspace{0.8mm} fb^{-1}$ and
$\sqrt{s}=3000\hspace{0.8mm}GeV$. Central panel: ${\cal L}=100, 500, 1500\hspace{0.8mm} fb^{-1}$
and $\sqrt{s}=1500\hspace{0.8mm}GeV$. Bottom panel: ${\cal L}=10, 100, 500\hspace{0.8mm} fb^{-1}$
and $\sqrt{s}=380\hspace{0.8mm}GeV$.}
\label{Fig.6}
\end{figure}

\begin{figure}[H]
\centerline{\scalebox{1}{\includegraphics{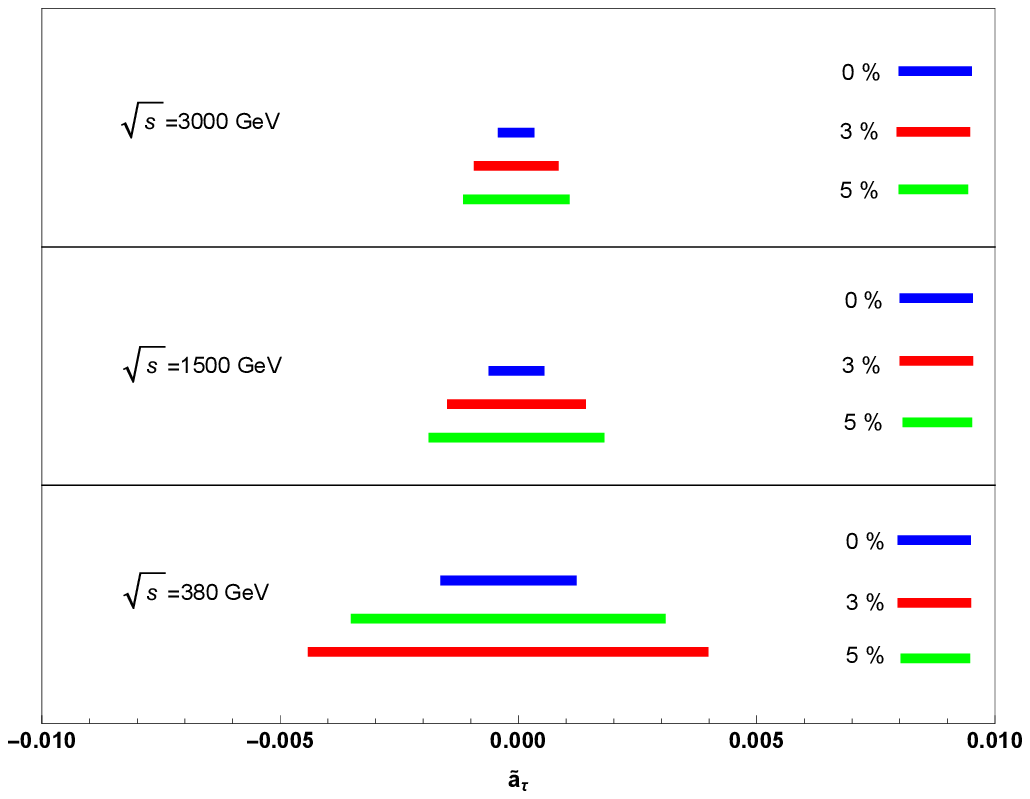}}}
\caption{ \label{fig:gamma15} Same as in Fig. 23, but for,
Top panel: $\delta_{sys}=0, 3, 5\hspace{0.8mm}\%$ with ${\cal L}=3000\hspace{0.8mm} fb^{-1}$ and $\sqrt{s}=3000\hspace{0.8mm}GeV$.
Central panel: $\delta_{sys}=0, 3, 5\hspace{0.8mm}\%$ with ${\cal L}= 1500\hspace{0.8mm} fb^{-1}$ and $\sqrt{s}=1500\hspace{0.8mm}GeV$.
Bottom panel: $\delta_{sys}=0, 3, 5\hspace{0.8mm}\%$ with ${\cal L}=500\hspace{0.8mm} fb^{-1}$ and $\sqrt{s}=380\hspace{0.8mm}GeV$.}
\label{Fig.6}
\end{figure}

\begin{figure}[H]
\centerline{\scalebox{1.3}{\includegraphics{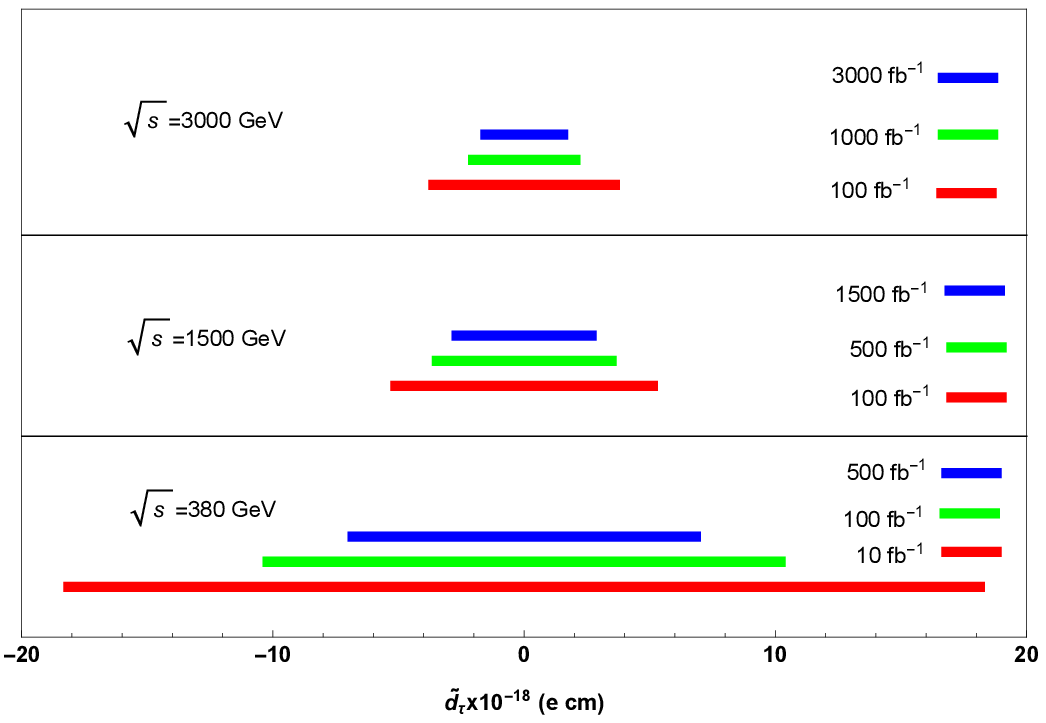}}}
\caption{ \label{fig:gamma15} Same as in Fig. 23, but for $\tilde{d}_\tau$.}
\label{Fig.6}
\end{figure}

\begin{figure}[H]
\centerline{\scalebox{1.1}{\includegraphics{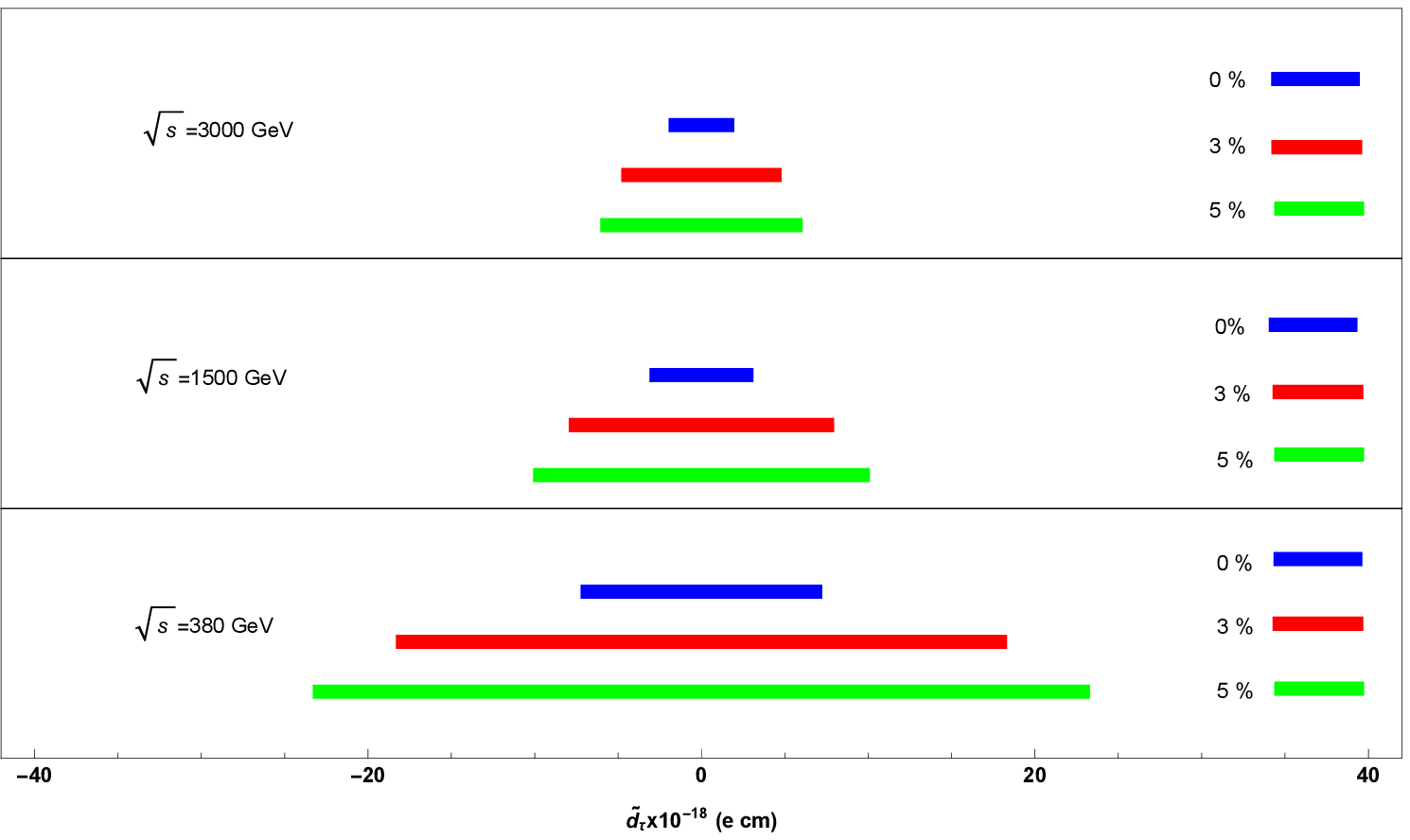}}}
\caption{ \label{fig:gamma15} Same as in Fig. 24, but for $\tilde{d}_\tau$.}
\label{Fig.6}
\end{figure}

\end{document}